\documentstyle[12pt,wrapfig2,epsf,citesort]{article}
\begin{document}
% Equation numbers for each sections.
\makeatletter
\renewcommand{\theequation}{\thesection.\arabic{equation}}
\@addtoreset{equation}{section}
\makeatother
\baselineskip=18pt plus 0.2pt minus 0.1pt
\parskip = 6pt plus 2pt minus 1pt
\newcommand{\lsim}{\,\lower0.9ex\hbox{$\stackrel{\displaystyle >}{\sim}$}\,}
\newcommand{\reseteqnum}{\setcounter{equation}{0}}
\newcommand{\sD}{D\hspace{-0.63em}/}
\newcommand{\al}{\alpha}
\newcommand{\be}{\beta}
\newcommand{\ve}{\varepsilon}
\newcommand{\ep}{\epsilon}
\newcommand{\la}{\lambda}
\newcommand{\La}{\Lambda}
\newcommand{\ga}{\gamma}
\newcommand{\si}{\sigma}
\newcommand{\de}{\delta}
\newcommand{\bx}{\mbox{\boldmath $x$}}
\newcommand{\bX}{\mbox{\boldmath $X$}}
\newcommand{\bY}{\mbox{\boldmath $Y$}}
\newcommand{\D}{\partial}
\newcommand{\rhovev}{(\rho v)^2}
\newcommand{\mr}{\frac{g^2}{\lambda}}
\begin{titlepage}
\title{\vskip -1.5cm
%\parbox{4cm}{\epsfxsize=3cm \epsfbox{kyoto-u.eps}}
\hfill
\parbox{4cm}{\normalsize KUCP-0084\\KUNS-1369\\HE(TH)~95/19
\\hep-th/9512064\\}\\
\vspace{1cm}
Valley Instanton versus Constrained Instanton
}
\author{Hideaki Aoyama\thanks{e-mail address:
\tt aoyama@phys.h.kyoto-u.ac.jp}
\\
{\normalsize\sl Department of Fundamental Sciences,}
\\
{\normalsize\sl Faculty of Integrated Human Studies,
Kyoto University, Kyoto 606-01, Japan}
\vspace{0.5cm}
\\
Toshiyuki Harano\thanks{e-mail address:
\tt harano@gauge.scphys.kyoto-u.ac.jp}
\\
{\normalsize\sl Department of Physics, Faculty of Science,
Kyoto University, Kyoto 606-01, Japan}
\vspace{0.5cm}
\\
Masatoshi Sato\thanks{e-mail address: \tt msato@gauge.scphys.kyoto-u.ac.jp}
\\
{\normalsize\sl Department of Physics, Faculty of Science,
Kyoto University, Kyoto 606-01, Japan}
\vspace{0.5cm}
\\
Shinya Wada\thanks{e-mail address: \tt shinya@phys.h.kyoto-u.ac.jp}
\\
{\normalsize\sl Graduate School of Human and Environmental Studies,
Kyoto University}\\
{\normalsize\sl Kyoto 606-01, Japan}}
\date{\normalsize December 1995}
\maketitle
\thispagestyle{empty}

\vspace*{-1cm}
\begin{abstract}
\normalsize
Based on the new valley equation, we propose the most plausible method
for constructing instanton-like configurations in the theory where the
presence of a mass scale prevents the existence of the classical
solution with a finite radius.
We call the resulting instanton-like configuration as valley instanton.
The detail comparison between the valley instanton and the constrained
instanton in $\phi^4$ theory and the gauge-Higgs system are carried out.
For instanton-like configurations with large radii, there appear remarkable
differences between them.
These differences are essential in calculating the baryon number
violating processes with multi bosons.
\end{abstract}
\end{titlepage}
\reseteqnum
\section{Introduction}
Quantum tunneling plays crucial roles in various
aspects of the quantum field theories.
To name a few in high energy physics,
supersymmetry breaking \cite{NSVZ,ADS,Yung,AKMRV},
baryon and lepton number violation phenomena in the standard
model \cite{thooft,manton,gold,Ring,Esp,AK},
asymptotic estimates of the perturbation theories
\cite{asym,brezin,parisi,fy} are among them.
Analysis of the phenomena is often carried out
in the imaginary-time path-integral formalism,
in which existence of the dominating configuration,
instantons or bounces, makes analytical treatment possible.

There however are cases when no
(regular) solution of the equation of motion exists.
A gauge theory with Higgs scalars is the most infamous,
although a class of scalar theories that exhibit similar features exists.
Not all the hope is lost, however.
The ordinary instanton does not exist in these theories
due to a scaling property:
It can be shown that among all the configurations of finite Euclidean
action, only the zero-radius configuration can
be a solution of equation of motion.
Consequently, all of the evaluations mentioned above were
done by the so-called constrained instanton formalism \cite{Aff}.
In this formalism, one introduces a constraint to
define a sub-functional space of finite radius configurations.
The field equation is solved in this subspace
and the finite-radius configuration similar to instantons
is defined. After doing the one-loop (and possibly higher order)
integral in this subspace,
the constraint parameter is integrated over.
The configuration constructed this way is called ``constrained
instanton''.
One problem about this method is that its validity
depends on the choice of the constraint:
Since in practice one does the Gaussian integration around the
solution under the constraint,
the degree of approximation depends on the way constraint is introduced.
Unfortunately, no known criterion guarantees the
effectiveness of the approximation.

This situation could be remedied once one realizes that
what we have near the point-like (true) instanton is the
valley \cite{AHSWlett}.
That is, although the zero-size instanton may be the dominating
configurations, it is expected to be followed by a series of
configurations that makes the valley of the action.
Therefore, instead of trying to cover all the neighborhoods of
the zero-radius instanton, one may cover the valley region,
which is expected to dominate the path-integral.
The trajectory along the valley bottom should correspond to
the scaling parameter, or the radius parameter of the instanton.
As such, the finite-size instanton can be defined as
configurations along the valley trajectory.
This is similar to a calculation in the electroweak theory in which
one evaluates the contribution of the instanton-anti-instanton
valley \cite{Yung,balyun,KR}.
Thus treating a single instanton as a configuration on the valley
provides a means of unifying the approximation schemes.
These configurations are named ``valley instanton''.

One convenient way to define the valley trajectory
is to use the new valley method \cite{silv,AKnv,aw}.
In this paper, we apply this formalism to construct the
actual valley instantons in the scalar system and the
gauge-Higgs system and investigate their features.
Since the constrained instanton has been used extensively in
the existing literature, our main purpose here is to establish
the existence of the valley instanton on a firm basis and
compare its properties with that of the constrained instanton.
This should serve as a starting point for the reanalysis of
the existing theories and results under the new light.

In the next section, we review the definition and the properties
of the new valley method.  Emphasis is placed on its
various advantages as a generalization of the
collective coordinate method. These features are
demonstrated in a toy model with two degrees of freedom.
In section \ref{sec:sft}, we study a scalar theory
where a mass scale prevents the existence of the finite-size instantons.
The valley instanton is constructed both analytically
and numerically and is compared with the constrained instanton.
More interesting and practical application of the idea
of the valley instanton is in the gauge-Higgs system.
The analysis of this system is carried
out in the following section along the similar line.
The last section gives the discussion and comments.

\section{New Valley Method}\label{sec:NV}

\def\shiki#1#2{\begin{eqnarray} #1 \label{eq:#2}
        \end{eqnarray}}
\def\eqa#1{(\ref{eq:#1})}

\def\lamin{\lambda}

The new valley method is most easily examined
in the context of the discretized theory.
The results thus obtained can be readily generalized to the continuum case.
We discretize the space-time and denote the resulting real bosonic variables
by $\phi_i$. These variables are the ``coordinates'' of the
functional space we carry out our analysis.
The bosonic action is written as a function of these variables;
$S = S(\phi_i).$
In this notation, the equation of motion is written as
$\partial_i S =0$, where
$\partial_j \equiv \partial / \partial \phi_j$.
We assume that the metric of the functional space is
trivial in these variables.  Otherwise, the metric should be
inserted in the following equations in a straightforward manner.

The new valley equation is the following;
\shiki{
D_{ij} \partial_j S = \lamin \partial_i S,
}{nvmdef}
where summation over the repeated indices are assumed
and
\shiki{
D_{ij} \equiv \partial_i\partial_j S.
}{ddef}
Since \eqa{nvmdef} has a parameter $\lamin$, it defines a
one-dimensional trajectory in the $\phi$-space.
The solution of the equation of motion apparently
satisfies the new valley equation \eqa{nvmdef}.
In this sense the new valley equation is an
extension of the (field) equation of motion.

According to the new valley equation \eqa{nvmdef},
the parameter $\lamin$ in the right-hand side of
\eqa{nvmdef} is one of the eigenvalues of the matrix $D_{ij}$.
Therefore the new valley equation specifies that the
``gradient vector'' $\partial_j S$ be parallel to the
eigenvector of $D_{ij}$ with the eigenvalue $\lamin$.
A question arises which eigenvalue of $D_{ij}$ should
we choose for \eqa{nvmdef}.
As we will show later, the new valley method
converts the eigenvalue $\lambda$ to a collective coordinate,
by completely removing $\lambda$ from the Gaussian integration and
introducing the valley trajectory parameter instead.
Thus the question is which eigenvalue ought to be
converted to a collective coordinate for the given theory.
In the scalar field theory with a false vacuum,
it was chosen to be the lowest eigenvalue, which corresponds
to the radius of the bounce solution \cite{aw}.
It was the negative eigenvalue near the bounce solution.
This was because the particle-induced false vacuum decay was
the subject of interest, for which smaller size bubbles are
relevant.
The general guideline, however, is to choose $\lambda$ to be the
eigenvalue with the smallest absolute value, {\it i.e.,} the
pseudo-zero mode, for the Gaussian integration for such direction
converges badly or diverges.  If a lower and negative
eigenvalue exists below $\lambda$, it simply creates the
imaginary part.
In the later sections, we follow this guideline.

The new valley equation \eqa{nvmdef} can be interpreted within a framework
of the variational method: Let us rewrite it as the following;
\shiki{
\partial_i \left(
{1 \over 2}\left( \partial_j S \right)^2 - \lamin S
\right) =0.
}{lagra}
This means that the norm of the gradient vector is extremized
under the constraint $S =$ const, where $\lamin$ plays the
role of the Lagrange multiplier.
In addition, we require that the norm be {\sl minimized}.
We are therefore defining the valley to be the trajectory
that is tangent to the most gentle direction,
which is a plausible definition, as we will see later in a simple example.
This also gives us an alternative explanation for the
fact that the solutions of the equation \eqa{nvmdef} form a trajectory.

\begin{wrapfigure}{c}{14cm}
\centerline{
\epsfxsize=10cm
\epsfbox{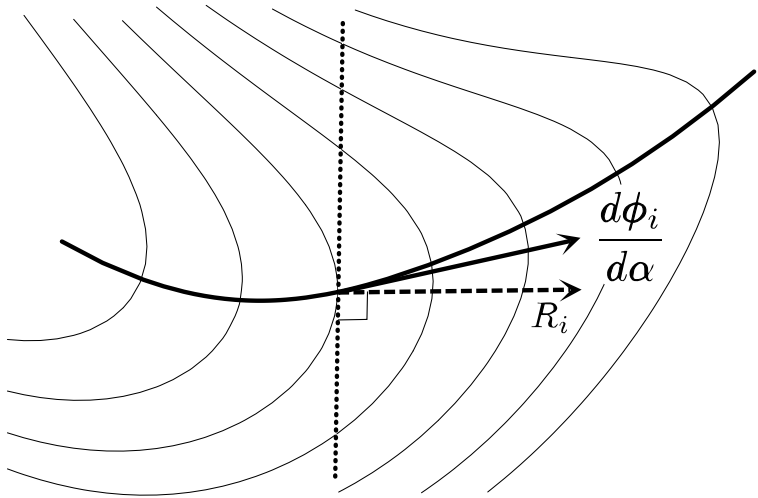}
}
\caption{A two-dimensional model of the functional space.
The thin solid lines denote the contours of the action $S$.
The thick solid line is the valley line. The direction of the
gradient vector $R_i$ and the tangent vector
$d \phi_i / d\alpha$ are denoted by the arrows.
The vertical dotted line is the subspace for the
one-loop integral.}
\label{fig:valillust}
\end{wrapfigure}

The functional integral is carried out along this valley line
in the following manner:
Let us parametrize the valley line by a parameter $\alpha$;
the solutions of \eqa{nvmdef} are denoted as $\phi(\alpha)$.
The integration over $\alpha$ is to be carried out exactly,
while for other directions the one-loop (or higher order)
approximation is applied.
We are to change the integration variables
($\phi_i$) to $\alpha$ and a subspace of $(\phi_i)$.
This subspace is determined uniquely by the following argument:
In expanding the action around $\phi(\alpha)$,
\shiki{
S(\phi) = S(\phi(\alpha)) + \partial_i S (\phi(\alpha))
\Bigl(\phi_i - \phi_i(\alpha)\Bigr)
+ {1 \over 2}D_{ij}
\Bigl(\phi_i - \phi_i(\alpha)\Bigr)\Bigl(\phi_j - \phi_j(\alpha)\Bigr)
+ \cdots ,}{sexpand}
the first order derivative term is {\sl not} equal to zero,
for $\phi(\alpha)$ is not a solution of the equation of motion.
In such a case, the $\hbar$-expansion is no longer the loop
expansion, as the tree contribution floods the expansion.
Therefore, we force this term to vanish by the choice of the subspace.
This choice of subspace is most conveniently
done by the Faddeev-Popov technique.
We introduce the FP determinant $\Delta(\phi(\alpha))$ by
the following identity;
\shiki{
\int d\alpha\ \delta\Bigl((\phi_i - \phi_i(\alpha))R_i \Bigr)
\Delta(\phi(\alpha))=1,
}{fpdef}
where $R_i (\phi(\alpha)) \equiv \partial_i S / \sqrt{ (\partial S)^2}$.
The solution to the above equation is,
\shiki{
\Delta(\phi(\alpha)) = \left| {d\phi_i(\alpha) \over d\alpha}
\Bigl\{R_i - \partial_i R_j (\phi_j - \phi_j(\alpha))\Bigr\}
\right| .
}{dfres}
In the one-loop approximation, the second term can be neglected
and the FP determinant contains the cosine of the angle between
the gradient vector and the vector tangent to the trajectory.
This is simply a Jacobian factor, because
the trajectory is not necessarily orthogonal to the chosen subspace.
This situation is illustrated in Fig.\ref{fig:valillust}.

Substituting \eqa{fpdef} into the path integral for the vacuum-to-vacuum
transition amplitude,
\shiki{
Z = {\cal N}\int \prod_j d\phi_j \ e^{-S},
}{pathint}
we obtain
\shiki{Z = {\cal N}\int d\alpha
\int\prod_j d\phi_j \delta\Bigl((\phi_i - \phi_i(\alpha))R_i \Bigr)
\left|{d\phi_i \over d\alpha} R_i \right|
e^{-S(\phi)}.}
{fppathint}
The second (functional) integration,
$\int\prod_j d\phi_j \delta((\phi_i - \phi_i(\alpha))R_i)$,
corresponds to the integration in the subspace denoted by
the dotted line in Fig.\ref{fig:valillust}.
At the one-loop order, the $\phi_i$ integration
yields $\det^\prime$, the determinant of $D_{ij}$ restricted to the subspace.
This subspace is orthogonal to the gradient vector,
which is the eigenvector with
the smallest eigenvalue $\lamin$ according to the new valley equation
\eqa{nvmdef}.
Therefore this subspace is the whole space less the direction of
the smallest eigenvalue of $D_{ij}$.
Therefore the resulting determinant is simply the ordinary determinant less
the smallest eigenvalue $\lamin$;
\shiki{{\rm det}^\prime = {\det \over \lamin}_{\displaystyle .}} {detprime}
We thus obtain the following expression at the one-loop order;
\shiki{
Z = {\cal N}^\prime
\int d \alpha \left|{d\phi_i \over d\alpha} R_i  \right|
{1\over\sqrt{\det^\prime}} \ e^{-S(\phi(\alpha))}.}
{zreduct}
In this sense, the new valley method converts the
smallest eigenvalue to the collective coordinate.
This exact conversion is quite ideal for the actual calculation
in the following sense:
In physical situation one often suffers from a negative,
zero, or positive but very small eigenvalue, which renders the
ordinary Gaussian integration meaningless or unreliable.
The new valley method saves this situation by converting
the unwanted eigenvalue to the collective coordinate.
The factor $\det^\prime$ is exactly free from this eigenvalue.
An added bonus to this property is that the resulting det$^\prime$
is quite easy to calculate; we simply calculate the whole determinant
and divide it by the smallest eigenvalue.
If several unwanted eigenvalues exist,
the new valley method can be extended
straightforwardly.  The equation should then specify that
the gradient vector $\partial_i S$ lie in the subspace
of the unwanted eigenvalues. This leads to a multi-dimensional
valley with all the advantages noted above.

There is another valley method, called ``streamline method" \cite{balyun},
which has been extensively used in the literatures.
It proposes to trace the steepest descent line starting
from a region of larger action.
By this definition, its Jacobian is trivial
at the one-loop order.
Nevertheless, its subspace for the perturbative calculation
has no relation to the eigenvalues of $D_{ij}$.
Therefore, the determinant is not
guaranteed to be free from the unwanted eigenvalue(s).
Another problem is that it is a flow equation in
the functional space:
Since it is not a local definition in the functional space,
it does not define any field equation.
Therefore the construction of the configurations
on the valley trajectory is quite difficult.
Another problem is that it suffers from instability if the valley is traced
from the {\sl bottom} of the valley.
This is not an issue for the problem
of the instanton and anti-instanton valley, for which
the higher end of the valley is known to be
the pair separated by infinite distance.
Yet this instability makes the streamline method useless
for the current problem,
for we only know the bottom of the valley,
the zero-size instanton.
For more detailed comparison of these methods
and other features of the new valley method,
we refer the readers to Ref.\cite{AKnv}.

\begin{wrapfigure}{c}{14cm}
\centerline{
\epsfxsize=8cm
\epsfbox{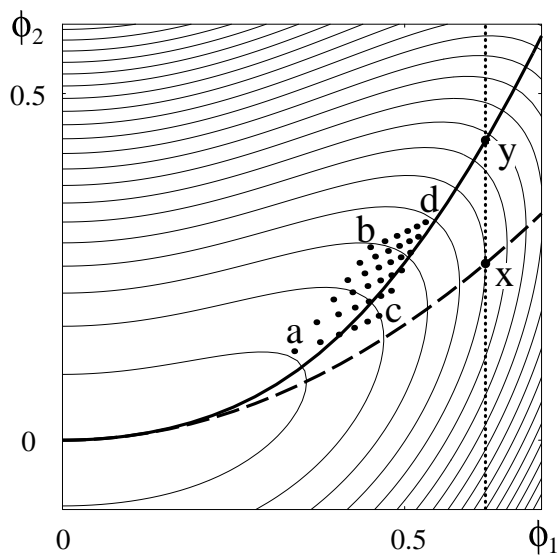}
}
\caption{The valley instanton and the constrained instanton in the toy
model.
The solid line denotes the valley instanton and the dashed line
shows the constrained instanton.
The dots show the positions of the saddle points of
the integrand of $O(p,q)$.}
\label{fig:toymodel}
\end{wrapfigure}

Let us discuss a toy model in a two-dimensional functional space,
to show the power of the new valley method.
The two degrees of freedom in the model is denoted
as $\phi_i$ ($i = 1,2$),
and the action is given by the following,
\shiki{
S(\phi_1, \phi_2) = {1 \over g^2}
\left(
(\phi_1^2 + \phi_2)^2 + 5(\phi_1^2 - \phi_2)^2
\right)_{\displaystyle .}
}{twoaction}
This is constructed by distorting a simple parabolic
potential so that the valley trajectory is not trivial.
The new valley equation is now a simple algebraic equation,
which can be solved numerically.
Alternatively, we could reparametrize the functional space
by $(r, \theta)$ defined by the following;
\shiki{
\phi_1^2 + \phi_2 = r \cos\theta, \hskip 3mm
\phi_1^2 - \phi_2 = {1 \over \sqrt{5}}r \sin\theta.
}{repra}
Under this parametrization $S = r^2/ g^2$.
Therefore following the variational interpretation of
the new valley equation, we can minimize the square of the norm of the
gradient vector, $(\partial_i S)^2$, as a function of $\theta$.
In Fig.\ref{fig:toymodel}, the thin lines denote the
$S=$ constant lines, and the thick solid line denotes
the new valley trajectory.

As an analogue of the constrained instanton formalism
in this toy model, we introduce a simple constraint.
Near the origin the valley trajectory extends to the $\phi_1$ direction.
Any reasonable constraint has to reproduce this property.
Therefore, as a simple example for the constraint,
we choose the following;
\shiki{\phi_1 = {\rm const.}}
{toyconst}
The solution of the constraint method is plotted in
Fig.\ref{fig:toymodel} by the dashed line.

In Fig.\ref{fig:toymodel}, it is apparent that the new valley trajectory
goes through the region of importance, while the constrained
trajectory does not.
This property can be displayed more explicitly
by considering the analogue of
the physical observables $O(p,q)$ defined by the following;
\shiki{
O(p,q) = \int_0^\infty d\phi_1 \int_{-\infty}^\infty d\phi_2
\ \phi_1^p \phi_2^q \ e^{- S(\phi_1, \phi_2)}.
}{opq}
Following the prescription given above, we have carried out
the numerical evaluation of $O(p,q)$
exactly and by the Gaussian (``one-loop")
approximation around the new valley trajectory and
the constrained trajectory.
Table.\ref{tab:toyratio} gives the ratio of
the one-loop values over the exact value.

As is seen in the table, the new valley method gives
a better approximation than the constraint method
in this range of $(p, q)$ consistently.
This could be explained in the following manner;
one could calculate the position of the maximum of
the integrand in \eqa{opq}, or the following
``effective action";
\shiki{\tilde S^{(p, q)}(\phi)
= S(\phi) - p \log \phi_1 - q \log \phi_2.}
{stilde}
The peak positions calculated in this manner
are denoted in Fig.\ref{fig:toymodel} by dots.
The point $a$ in the left bottom is for
$(p, q) = (2,2)$, $b$ for $(2,12)$, $c$ for $(12,2)$,
and $d$ for $(12,12)$.
The dots are distributed around the new valley trajectory.
This explains the fact that the new valley trajectory
gives the better approximation over the constrained
trajectory.
Only when the power of one of the observable gets high
enough as in the point $c$,
the constraint method starts to give a little better values.
However, this is exactly the range in which
the ordinary instanton calculation starts to fail;
For the $n$-point function with $n > O(1/g^2)$,
it is well known that one needs to take into account
the effect of the external particles on the instanton itself.
The power of the valley method lies in the fact that
even in this range the calculation is done with good
accuracy for generic type of observables, like the (12,12)
case. Only when the observable is very special, like
the (12,2) case, the constraint method tuned to
that operator may give reasonable estimates.

\begin{wraptable}{c}{14cm}
\begin{center}
\doublerulesep=0pt
\def\arraystretch{1.3}
\begin{tabular}{@{\vrule width 1.0pt \quad}c @{\quad} |
c|c|c|c|c|c @{\ \vrule width 1.0pt}}
\noalign{\hrule height 1.0pt}
$p\backslash q$ &2& 4 & $6$ & 8 &
10 & 12 \\
\hline
2  & 0.827  & 0.679  & 0.585  & 0.520  & 0.474  & 0.439 \\
   & 0.545  & 0.248  & 0.110  & 0.049  & 0.022  & 0.010 \\
\hline
4  & 0.997  & 0.870  & 0.767  & 0.689  & 0.629  & 0.581 \\
   & 0.667  & 0.324  & 0.149  & 0.068  & 0.030  & 0.014 \\
\hline
6  & 1.061  & 0.986  & 0.900  & 0.826  & 0.764  & 0.712 \\
   & 0.737  & 0.384  & 0.185  & 0.086  & 0.040  & 0.018 \\
\hline
8  & 1.064  & 1.046  & 0.991  & 0.931  & 0.876  & 0.826 \\
   & 0.783  & 0.432  & 0.218  & 0.105  & 0.050  & 0.023 \\
\hline
10 & 1.029  & 1.063  & 1.041  & 1.003  & 0.960  & 0.918 \\
   & 0.815  & 0.473  & 0.248  & 0.123  & 0.060  & 0.028 \\
\hline
12 & 0.971  & 1.046  & 1.058  & 1.043  & 1.017  & 0.985 \\
   & 0.839  & 0.508  & 0.276  & 0.141  & 0.070  & 0.034 \\
\noalign{\hrule height 1pt}
\end{tabular}
\end{center}
\caption{The ratio of the $O(p,q)$ estimated by
new valley method (the upper column) and the constraint method
(the lower column) over the exact value for $g=0.2$. }
\label{tab:toyratio}
\end{wraptable}

Let us note a tricky point in comparing the valley instanton
and the constrained instanton.
If we compare them with the same parameter values,
the constrained instanton has smaller action than the
valley instanton.  This is not a contradiction, nor it means
the constrained instanton has a larger contribution:
This becomes clear by considering the dotted vertical line
in Fig.\ref{fig:toymodel}.
The point $x$ and $y$ are the valley and the constrained configurations,
respectively, which have the same value of $\phi_1$.
The configuration $y$ has a larger action than $x$,
by the definition of the constraint method.
However, as we have seen above, this does not mean
that the constrained trajectory gives a better approximation.
(One way to illustrate this point is that if we compare
the configurations at the same distance, $\int d\alpha
|(d\phi_i / d\alpha) R_i |$, from the origin, the valley configuration has
the smaller action.)
The same situation will be seen in the results of the following
sections; when we compare the instantons at the same value of the
constraint parameter, the constrained instanton has a smaller action than
that of the valley instanton.
This is a red herring and has nothing to do with
the relevancy of the valley instanton.

In this analysis we took a very simple
constraint \eqa{toyconst}.
Alternatively we could take
some ad-hoc constraint as long as any of these
yield ``finite-size" instantons, {\it i.e.},
points away from the origin.
Most of these constraints yield essentially similar
results.
Only when the trajectory goes through the dotted area
in the Fig.\ref{fig:valillust}, one can obtain good quantitative
results.  Such a trajectory, however, is destined to be
very close to the new valley trajectory.
Therefore the new valley method, defined without any
room for adjustment and guaranteed to give good results
is superior to the constrained method.

Finally, we note that in the actual calculation of the solutions in the
continuum space-time it is useful to introduce an auxiliary field.
Let us denote all the real bosonic fields in the theory by  $\phi_\alpha(x)$.
We introduce the auxiliary field $F_\alpha (x)$ for each bosonic
field and write the new valley equation as follows;
\shiki{
\sum_\beta \int d^4 y
{\delta^2 S \over \delta\phi_{\alpha}(x) \delta\phi_{\beta}(y)}
F_{\beta}(y) = \lambda F_{\alpha}(x),
\hskip 4mm
F_\alpha (x) = {\delta S \over \delta\phi_{\alpha}(x)}_{\displaystyle .}
}{gennv}
This is a set of second-order differential equation,
which can be analyzed by the conventional methods.
The solution of the ordinary field equation of motion
is a solution of \eqa{gennv} with $F_\alpha (x) =0$.
In other words, $F_\alpha (x)$ specifies where and
how much the valley configuration
deviates from the solution of equation of motion.
This property is useful for the
qualitative discussion of the properties of the valley
configurations.
The analysis in the following sections will be carried out
in this auxiliary field formalism \eqa{gennv}.

\section{The scalar $\phi^4$ field theory}\label{sec:sft}
We consider the scalar $\phi_4^{4}$ theory with Euclidean
Lagrangian,
\begin{eqnarray}
{\cal L}=\frac{1}{g^2}\left[ \frac{1}{2}(\D_{\mu} \phi)^2 +\frac{1}{2}
\mu^2 \phi^2 -\frac{1}{4!}\phi^4 \right].
\label{scalarlagrangian}
\end{eqnarray}
The negative sign for the $\phi^4$ term is relevant for
performing the asymptotic estimate in the theory with a positive $\phi^4$
term \cite{asym,brezin,parisi,fy,Aff}.
This model allows finite-size instantons only for $\mu=0$;
otherwise, a simple scaling argument shows
that the action of any finite-size configuration
can be reduced by reducing its size.

\subsection{Valley instanton}
For this model, the new valley equation introduced in the previous section is
\begin{eqnarray}
\begin{array}{cc}
&\displaystyle -\partial_{\mu}\partial_{\mu} \phi + \mu^2 \phi
- \frac{1}{3!}\phi^3 =F,
\\[0.5cm]
&\displaystyle \left( -\partial_{\mu}\partial_{\mu}+\mu^2 -\frac{1}{2}
\phi^2\right)F=\lambda F.
\end{array}
\label{scalarge:gennv}
\end{eqnarray}
Now we introduce the scale parameter $\rho$
defined by $\phi(0) \equiv 4\sqrt{3}/\rho$, in order to
fix the radius of the instanton solution to be unity.
We rescale fields and variables as in the following;
\begin{equation}
r=\frac{\sqrt{x^2}}{\rho},\quad
\lambda=\mu^2\nu,\quad
\phi(x)=\frac{h(r)}{\rho},\quad
F(x)=\frac{\mu^2}{\rho} f(r).
\label{scalarrescale}
\end{equation}
The new valley equation (\ref{scalarge:gennv}) is given by the following
under this rescaling;
\begin{eqnarray}
&&-\frac{1}{r^3}\frac{d}{dr}\left(r^3\frac{dh}{dr}\right)
+(\rho\mu)^2h-\frac{1}{3!}h^3=(\rho\mu)^2f,
\label{scalarnewvalleyeq1}\\
&&-\frac{1}{r^3}\frac{d}{dr}\left(r^3\frac{df}{dr}\right)
+(\rho\mu)^2f-\frac{1}{2}h^2f=(\rho\mu)^2\nu f.
\label{scalarnewvalleyeq2}
\end{eqnarray}

This system has an instanton solution in the massless limit, $\rho \mu \to
0$ \cite{asym,brezin}.
In this limit, (\ref{scalarnewvalleyeq1}) reduces to the equation of
motion and (\ref{scalarnewvalleyeq2})
the equation for the zero-mode fluctuation, $f$, around the instanton solution.
The solution is the following;
\begin{eqnarray}
\begin{array}{cc}
\displaystyle h_{0}=\frac{4\sqrt{3}}{1+r^{2}},
&\displaystyle f_{0}
=C\left(\frac{4\sqrt{3}}{1+r^2}-\frac{8\sqrt{3}}{(1+r^2)^2}\right),
\end{array}
\label{scalarinstanton}
\end{eqnarray}
where $C$ is an arbitrary constant.
Note that
solution $f_0$ is obtained from $\partial \phi_0 (x) /\partial \rho$,
$\phi_0=h_0/\rho$.

Let us construct the valley instanton in the scalar $\phi^4$ theory
analytically.
When $\rho \mu$ is very small but finite, the valley instanton
is expected to have $\rho \mu$ corrections to (\ref{scalarinstanton}).
On the other hand, at large distance from the core region of the
valley instanton, since the term of $O(h^2)$ is negligible, the
valley equation can be linearized.
This linearized equation can be solved easily.
By matching the solution near the core and the solution in the asymptotic
region in the overlapping intermediate region, we can construct
approximate solution analytically.
We will carry out this procedure in the following.

In the asymptotic region,  $(\rho\mu)^2\gg h^2$,
the linearized valley equation is the following;
\begin{eqnarray}
\begin{array}{cc}
&\displaystyle -\frac{1}{r^3}\frac{d}{dr}\left(r^3\frac{dh}{dr}\right)
+(\rho\mu)^2h=(\rho\mu)^2f,
\\[0.5cm]
&\displaystyle -\frac{1}{r^3}\frac{d}{dr}\left(r^3\frac{df}{dr}\right)
+(\rho\mu)^2f=(\rho\mu)^2\nu f.
\end{array}
\label{scalarlinernveq}
\end{eqnarray}
The solution of these equations is
\begin{equation}
h(r)=C_1G_{\rho\mu}(r)+\frac{f}{\nu},\quad
f(r)=C_2G_{\rho\mu\sqrt{1-\nu}}(r),
\label{scalarasymptoticsol}
\end{equation}
where $C_1$ and $C_2$ are arbitrary functions of $\rho \mu$.
The function $G_{m}(r)$ is
\begin{equation}
G_{m}(r)=\frac{m K_{1}(m r)}{(2\pi)^{2}r},
\end{equation}
where $K_{1}$ is a modified Bessel function.
The functions $f$ and $h$ decay exponentially at large $r$.
In the region of $r \ll (\rho \mu)^{-1}$, $r \ll (\rho \mu \sqrt{1-\nu})^
{-1}$, $f$ and $h$ can be expanded in series as the following;
\begin{eqnarray}
\begin{array}{lll}
&\displaystyle h=\frac{C_1}{(2\pi)^2}
\left[\frac{1}{r^2}+\frac{1}{2}(\rho\mu)^2\ln(\rho\mu rc)
+\cdots\right]\\[0.5cm]
&\displaystyle \hspace{1.7ex}+\frac{C_2}{(2\pi)^2\nu}
\left[\frac{1}{r^2}+\frac{1}{2}(\rho\mu)^2(1-\nu)
\ln(\rho\mu\sqrt{1-\nu}rc)+\cdots\right],
\\[0.5cm]
&\displaystyle f=\frac{C_2}{(2\pi)^2}
\left[\frac{1}{r^2}+\frac{1}{2}(\rho\mu)^2(1-\nu)
\ln(\rho\mu\sqrt{1-\nu}rc)+\cdots\right].
\end{array}
\label{scalarexpand}
\end{eqnarray}
In the above, $c= e^{\gamma-1/2}/2$, where $\gamma$ is
the Euler's constant.

Near the origin, we expect that the valley
instanton is similar to the ordinary instanton.
It is convenient to define $\hat{h}$ and $\hat{f}$ as the following;
\begin{equation}
h=h_0+(\rho\mu)^2\hat{h},\qquad
f=f_0+(\rho\mu)^2\hat{f},
\label{scalarperturbation}
\end{equation}
where $C$ in $f_0$ is the function of $\rho \mu$ and is decided in
the following.
The ``core region'' is defined as
\begin{equation}
h_0\gg (\rho\mu)^2\hat{h},\qquad
f_0\gg (\rho\mu)^2\hat{f}.
\label{scalarassumption2}
\end{equation}
The valley equation for perturbation field $\hat{h}$, $\hat{f}$ becomes
\begin{eqnarray}
\begin{array}{ll}
&\displaystyle -\frac{1}{r^3}\frac{d}{dr}\left(r^3\frac{d\hat{h}}{dr}\right)
-\frac{1}{2}h_0^2\hat{h}=f_0-h_0,
\\[0.5cm]
&\displaystyle -\frac{1}{r^3}\frac{d}{dr}\left(r^3\frac{d\hat{f}}{dr}\right)
-\frac{1}{2}h_0^2\hat{f}=(\nu-1)f_0+h_0f_0\hat{h}.
\end{array}
\label{scalarasymptoticveq}
\end{eqnarray}
The left-hand side of (\ref{scalarasymptoticveq}) has a zero mode,
$\varphi$. It satisfies the equation,
\begin{eqnarray}
-\frac{1}{r^3} \frac{d}{dr}\left(r^3 \frac{d \varphi}{d r}\right)
- \frac{1}{2} h_0^2 \, \varphi =0,
\label{scalarzeromodeeq}
\end{eqnarray}
and is given by the following;
\begin{equation}
\varphi=\frac{4\sqrt{3}}{1+r^2}-\frac{8\sqrt{3}}{(1+r^2)^2}.
\label{scalarzeromode}
\end{equation}
We multiply $r^3 \varphi$ to both sides of (\ref{scalarasymptoticveq}),
and integrate them from $0$ to $r$.
The existence of zero mode $\varphi$ makes it possible
to integrate the left-hand side of (\ref{scalarasymptoticveq}).
As a result of the integration by parts, only the surface terms remain
and we obtain,
\begin{eqnarray}
& \displaystyle -r^3\varphi\frac{d\hat{h}}{dr}+r^3\frac{d\varphi}{dr}\hat{h}
=\int_0^{r}dr'r'^3\varphi\left[f_0-h_0\right],
\label{valleyeqpart21}
\\[0.5cm]
& \displaystyle -r^3\varphi\frac{d\hat{f}}{dr}+r^3\frac{d\varphi}{dr}\hat{f}
=\int_0^{r}dr'r'^3\varphi\left[(\nu-1)f_0+h_0f_0\hat{h}\right].
\label{valleyeqpart22}
\end{eqnarray}
First, using (\ref{scalarinstanton}) and (\ref{scalarzeromode}),
we can find that the right-hand side of (\ref{valleyeqpart21}) is
proportional to $\ln r$ at $r \gg 1$.
Thus (\ref{valleyeqpart21}) becomes
\begin{eqnarray}
&\displaystyle r \frac{d \hat{h}}{dr}+2 \hat{h}
 =4\sqrt{3}\,(1-C) \ln r.
\label{valleyeqpart31}
\end{eqnarray}
This equation can be solved easily at $r\gg 1$.
Using the solution of (\ref{valleyeqpart31}), we can find the
right-hand side of (\ref{valleyeqpart21}) also proportional to $\ln r$ as
the following;
\begin{eqnarray}
&\displaystyle  r \frac{d \hat{f}}{dr}+2 \hat{f}
 =4\sqrt{3}\,C (1-\nu) \ln r.
\label{valleyeqpart32}
\end{eqnarray}
Finally, we obtain $\hat{h}$ and $\hat{f}$ at $r\gg 1$,
\begin{equation}
 \hat{h}=2\sqrt{3}\, (1-C)\ln r+\cdots, \quad
 \hat{f}=(1-\nu)2\sqrt{3}\, C\ln r+\cdots.
\end{equation}
For these solutions to meet to (\ref{scalarexpand}),
the parameters need to be the following;
\begin{equation}
C_1=0,\quad
C_2=4\sqrt{3}\,(2\pi)^2,\quad
\nu=1,\quad
C=1,
\label{scalarmatchingpara}
\end{equation}
as $\rho \mu =0$.
Now we have obtained the solution of the new valley equation;
\begin{equation}
h(r)=\left\{
    \begin{array}{ll}
\displaystyle      \frac{4\sqrt{3}}{1+r^2},&
\quad  \mbox{if} \quad r \ll (\rho \mu)^{-1/2};  \\[0.4cm]
\displaystyle      \frac{4\sqrt{3}}{r^2} + o\left((\rho \mu)^2 \right),&
\quad \mbox{if} \quad (\rho \mu)^{-1/2} \ll r \ll (\rho \mu)^{-1}; \\[0.4cm]
\displaystyle      4 \sqrt{3}\, (2 \pi^2) \,G_{\rho \mu \sqrt{1-\nu}}\,(r),&
\quad \mbox{if} \quad (\rho \mu)^{-1/2} \ll r.
     \end{array} \right.
\label{scalarinstanalyeq}
\end{equation}

Let us discuss the consistency of our analysis.
In the construction of the analytical solution, especially
in the argument of the matching of the core and asymptotic region solution,
we have implicitly assumed that there exists an overlapping region
where both (\ref{scalarlinernveq}) and
(\ref{scalarasymptoticveq}) are valid.
Using the solution (\ref{scalarinstanalyeq}), it is found that
(\ref{scalarlinernveq}) is valid in the
region of $r \gg (\rho \mu)^{-1/2}$, and
(\ref{scalarasymptoticveq}) is valid
in the region of $r \ll (\rho \mu)^{-1}$.
Therefore in the above analysis we have limited
our calculation in
the overlapping region $(\rho \mu)^{-1/2} \ll r \ll (\rho \mu)^{-1}$.

We calculate the action of the valley instanton using the above solution.
Rewriting the action in terms of $h(r)$, we find
\begin{equation}
S=\frac{\pi^2}{g^2}\int_0^\infty dr\, r^3 \left[
\left(\frac{dh}{dr}\right)^2 + (\rho \mu)^2 h^2 - \frac{1}{12} h^4 \right].
\end{equation}
Substituting the analytic solution for $S$, we obtain
\begin{equation}
S=\frac{16 \pi^2}{g^2} + O\left((\rho \mu)^2\right).
\label{scalarresultactionnv}
\end{equation}
The leading contribution term comes from the ordinary instanton
solution, and the correction term comes from the distortion of
the instanton solution.

\subsection{Constrained instanton}
In this subsection, we consider the constrained instanton in the scalar
$\phi^4$ theory, following the construction in Ref.\cite{Aff}.
We require the constraint in the path integral.
The field equation under the constraint is
\begin{eqnarray}
\frac{\delta S}{\delta \phi} + \sigma \frac{\delta O}{\delta \phi}=0,
\end{eqnarray}
where $\sigma$ is a Lagrange multiplier.
The functional $O$ had to satisfy the certain scaling properties
that guarantee the existence of the solution \cite{Aff}.
We choose it as follows;
\begin{eqnarray}
O=\int d^4 x \,\frac{\phi^6}{6}.
\end{eqnarray}
This choice is one of the simplest for constructing the constrained
instanton in the scalar $\phi^4$ theory.
Again, adopting the rescaling (\ref{scalarrescale}),
the equation of motion under this constraint becomes
\begin{eqnarray}
-\frac{1}{r^3}\frac{d}{dr}\left(r^3\frac{dh}{dr}\right)
+(\rho\mu)^2h-\frac{1}{3!}h^3+(\rho\mu)^2\tilde{\sigma}h^5=0,
\label{scalarconstrainedeq}
\end{eqnarray}
where we rescale the parameter $\sigma$ as $\sigma=(\rho\mu)^2\widetilde
{\sigma}$.
The solution of this equation can be constructed in a manner similar to the
previous section.
To carry out the perturbation calculation in the core region,
we replace the field variable as $h=h_0 + (\rho \mu)^2 \hat{h}$, where
$h_0 \gg (\rho \mu)^2 \hat{h}$.
The field equation of $\hat{h}$ becomes
\begin{equation}
-\frac{1}{r^3}\frac{d}{dr}\left(r^3\frac{d\hat{h}}{dr}\right)
-\frac{1}{2}h_0^2\hat{h}=-h_0-\tilde{\sigma}h_0^5.
\label{scalarconsthhateq}
\end{equation}
We multiply this equation by the zero mode (\ref{scalarzeromode}) and
integrate this from $0$ to $r$.
Then we obtain
\begin{equation}
-r^3\varphi\frac{d\hat{h}}{dr}+r^3\frac{d\varphi}{dr}\hat{h}=
-\int_0^{r}dr'r'^3\varphi\left[h_0+\tilde{\sigma}h_0^5\right].
\end{equation}
The solution of this equation in the region where $r \gg 1$ is
\begin{equation}
\hat{h}=2\sqrt{3}\ln r -\frac{192\sqrt{3}}{7}\tilde{\sigma}
+\cdots.
\label{scalarconstcoresol}
\end{equation}
In the asymptotic region, we consider the field variable and the
parameter as $h^2 \ll (\rho \mu)^2$, $\widetilde{\sigma}h^4\ll 1$.
Under this condition, the field equation of the asymptotic region becomes
\begin{equation}
-\frac{1}{r^3}\frac{d}{dr}\left(r^3\frac{dh}{dr}\right)
+(\rho\mu)^2h=0.
\label{scalarconstfieldeqatasym}
\end{equation}
The solution is
\begin{equation}
h(r)=C_1G_{\rho\mu}(r),
\end{equation}
where $C_1$ is arbitrary constant.
In the region of $r \ll (\rho \mu)^{-1}$, this solution can be expanded as
 the following;
\begin{equation}
h=\frac{C_1}{(2\pi)^2}\left[\frac{1}{r^2}
+\frac{1}{2}(\rho\mu)^2\ln(\rho\mu rc)+\cdots\right].
\label{scalarconstasymsol}
\end{equation}
Matching this solution and the core region solution
(\ref{scalarconstcoresol}), parameters are determined as the following;
\begin{equation}
C_1=4\sqrt{3}(2\pi)^2,\quad \tilde{\sigma}=-\frac{7}{96}\ln(\rho\mu).
\label{scalarconstparameter}
\end{equation}

To summarize, the analytical solution of the constrained instanton
we have obtained is the following,
\begin{equation}
h(r)=\left\{
    \begin{array}{ll}
\displaystyle \frac{4\sqrt{3}}{1+r^2},&
\quad \mbox{if} \quad r \ll (\rho \mu)^{-1/2}; \\[0.4cm]
\displaystyle \frac{4\sqrt{3}}{r^2} + 2 \sqrt{3} (\rho \mu)^2 \ln(\rho \mu rc)
+ o\left((\rho \mu)^2 \right), &
\quad \mbox{if} \quad (\rho \mu)^{-1/2} \ll r \ll (\rho \mu)^{-1}; \\[0.4cm]
\displaystyle 4 \sqrt{3}\, (2 \pi^2) \,G_{\rho \mu }\,(r),&
\quad \mbox{if} \quad (\rho \mu)^{-1/2} \ll r.
     \end{array} \right.
\label{scalarconstanalyeq}
\end{equation}

The action of the constrained instanton is given by the following;
\begin{equation}
S=\frac{16 \pi^2}{g^2} - 96 \pi^2 (\rho \mu)^2 \ln (\rho \mu)
+ O\left((\rho \mu)^2\right).
\label{scalarresultaction}
\end{equation}
This differs from the action of the valley instanton at the next-to-leading
order.
This correction term shows that the constrained instanton is more distorted
from the ordinary instanton than the
valley instanton.

\subsection{Numerical analysis}
In this subsection, we calculate the valley equation
(\ref{scalarnewvalleyeq1}), (\ref{scalarnewvalleyeq2}) and
the constrained equation (\ref{scalarconstrainedeq}) numerically.
Then we compare the valley and the constraint instanton.

Each of the equations is the second order differential equation, so
we require two boundary conditions for each field variable to
decide the solution.
We require all the field variables are regular at the origin.
The finiteness of the action requires $h, f\to 0$ faster than $1/r^2$ at
infinity.
In solving (\ref{scalarnewvalleyeq1}) and (\ref{scalarnewvalleyeq2}),
we adjust the parameter $\nu$ and $f(0)$ so that $h,f \to 0$ at infinity
for the fixed $\rho \mu$.
In the similar way, in case of the constrained instanton, the parameter
$\sigma$ is determined so that $h\to 0$ at infinity.

\begin{wrapfigure}{c}{14cm}
\centerline{
\epsfxsize=10cm
\epsfbox{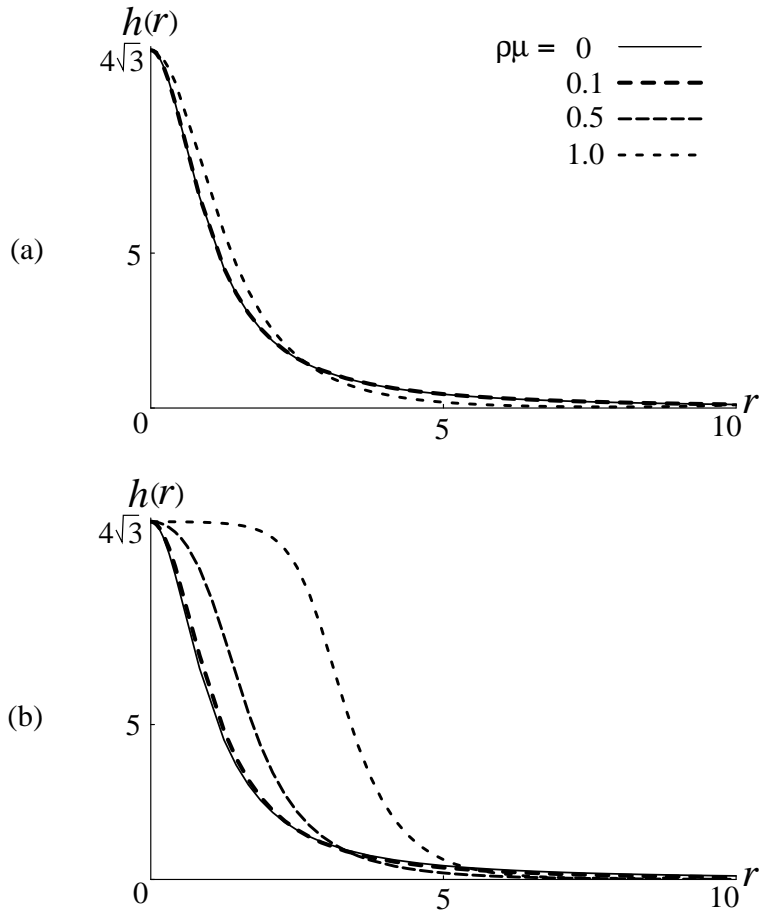}
}
\caption{(a) Shapes of the numerical solution of the valley
instanton,
 $h(r)$, for $\rho \mu=$$0,$ $0.1,$ and $1.0$ near the origin.
The solid line denotes the original instanton, $h_0$.
(b) Shapes of the constrained instanton for $\rho \mu=$$0,$ $0.1,$ $0.5,$
and $1.0$.}
\label{fig:scalarconfig}
\end{wrapfigure}

Numerical solutions of the valley equation near the origin are plotted
in Fig.\ref{fig:scalarconfig} (a) for $\rho \mu=0.1,\, 1.0$.
The solid line shows the instanton solutions (\ref{scalarinstanton}),
which corresponds to $\rho \mu=0$.
The numerical solutions of the constrained instanton are also plotted in
Fig.\ref{fig:scalarconfig} (b) for $\rho \mu=0.1, \,0.5,\,1.0$.
Both the valley and the constraint solution for $\rho \mu=0.1$ agree
with the analytical result.
As $\rho \mu$ becomes large, both solutions are deformed from the original
instanton solution.
We find that the distortion of the constrained instanton
is much larger than that of the valley instanton.
This also agrees with the analytic result.
In the analytical solution (\ref{scalarconstanalyeq}), the correction
term $2 \sqrt{3} (\rho \mu)^2 \ln(\rho \mu rc)$ contributes to this
distortion.
On the other hand, the correction term of the valley instanton
(\ref{scalarconstanalyeq}) is $o\left((\rho \mu)^2\right)$, which is
smaller than the previous one.
In addition, we find that the exponentially damping behavior
of the analytical solution in the asymptotic region, where
$r \gg (\rho \mu)^{-1}$ agrees with the result of numerical analysis.

\begin{wrapfigure}{c}{14cm}
\centerline{
\epsfxsize=12cm
\epsfbox{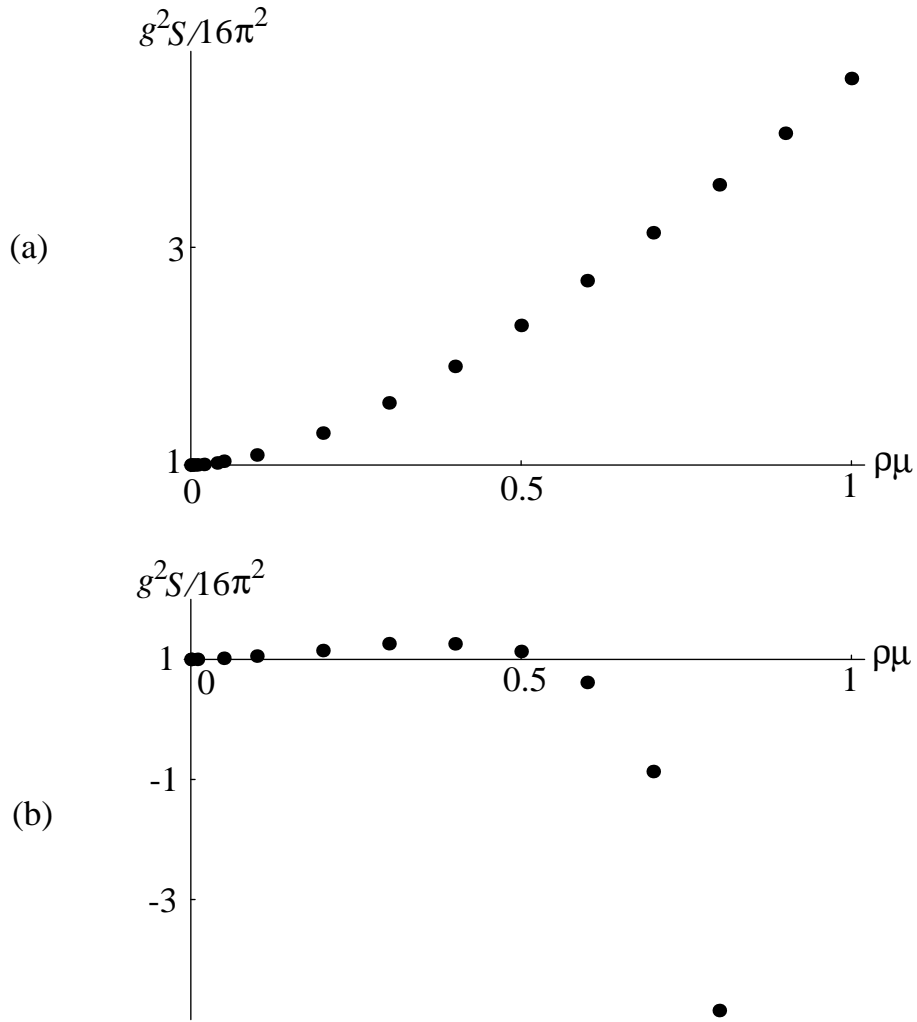}
}
\caption{(a) The action $S$ (in units of $g^2/16 \pi^2$) of the
numerical solution
of the valley equation as a function of the parameter $\rho \mu$.
(b) The action $S$ of the constrained instanton.}
\label{fig:scalaraction}
\end{wrapfigure}

The values of the action of the valley instanton for $\rho \mu=$$0.0001 \sim
1.0$ are plotted in Fig.\ref{fig:scalaraction} (a).
If $\rho \mu$ is very small, this result is consistent with
(\ref{scalarresultactionnv}).
The values of the action of the constrained instanton for $\rho \mu=$
$0.0001 \sim 0.8$ are plotted in Fig.\ref{fig:scalaraction} (b).
This figure shows that the behavior of the action is similar to that of
the valley instanton when $\rho \mu$ is very small.
When  $\rho \mu$ becomes large, the behavior of the action
is different from the valley instanton case.

We summarize all the numerical data in the Table \ref{tab:sftresult}.

\eject{}~\vfill
\vfill\eject

\begin{wraptable}{c}{14cm}
\begin{center}
\doublerulesep=1pt
\def\arraystretch{1.3}
\begin{tabular}{@{\vrule width1.0pt\quad}c@{\quad\vrule width1.0pt \,
}c|c|c||c|c@{\ \vrule width1.0pt}}\noalign{\hrule height 1.0pt}
&\multicolumn{3}{c||}{\centering valley instanton}&
\multicolumn{2}{c@{\vrule width1.0pt}}{constrained instanton}\\
\noalign{\hrule height 1pt}
{$\rho \mu$ } & {$\nu$} & $f(0)$ & $g^{2}S/16\pi^2$ & {$\sigma$} &
$g^{2}S/16\pi^2$\\
\hline
0.0001 & 0.971  & -7.14 & 1.000002 & 0.000000079 & 1.000008\\
\hline
0.001 & 0.958 & -7.21 & 1.000021 & 0.000000394 & 1.000021\\
\hline
0.01 &  0.940 & -7.44 & 1.001585 & 0.000019342 & 1.001281\\
\hline
0.05 & 0.917  & -7.93 & 1.035008 & 0.000257518 & 1.019044\\
\hline
0.1 & 0.901  & -8.43 & 1.092489 & 0.000661283 & 1.055766\\
\hline
0.2 & 0.876  & -9.40 & 1.293706 & 0.001437910 & 1.148013\\
\hline
0.3 & 0.855  & -10.41 & 1.571073 & 0.002055520 &  1.262493\\
\hline
0.4 & 0.833  & -11.46 & 1.906408 & 0.002508380 & 1.259776\\
\hline
0.5 & 0.811  &  -12.57 & 2.282639 & 0.002817690 & 1.132618\\
\hline
0.6 & 0.787  & -13.71 & 2.693832 & 0.003008290 & 0.616666\\
\hline
0.7 & 0.762  & -14.87 & 3.134111 & 0.003104190 & -0.867477\\
\hline
0.8 &  0.736  & -16.01 & 3.574478 & 0.003127540 & -4.848880\\
\hline
0.9 &  0.709  & -17.12 & 4.047787 & 0.003098220 & -15.87237\\
\hline
1.0 & 0.682  & -18.18 & 4.550929 & 0.003033310 & -49.97363\\
\noalign{\hrule height 1.0pt}
\end{tabular}
\end{center}
\caption{The numerical data of the valley instanton and the constrained
instanton in the scalar theory.}
\label{tab:sftresult}
\end{wraptable}

\section{The gauge-Higgs system}\label{sec:GH}
We consider the SU(2) gauge theory with one scalar Higgs doublet,
which has the following action $S=S_g+S_h$;
\begin{eqnarray}
&&S_g=\frac{1}{2 g^2} \int d^4 x\ {\rm tr} F_{\mu \nu}  F_{\mu \nu},
\label{gaugeaction}
\\
&&S_h=\frac{1}{ \lambda}\int d^4 x\left\{\left(D_{\mu} H\right)^{\dagger}
\left(D_{\mu} H\right)+
\frac{1}{8}\left(H^{\dagger}H-v^2\right)^2\right\},
\label{higgsaction}
\end{eqnarray}
where $F_{\mu\nu}=\partial_{\mu} A_{\nu}-\partial_{\nu}
A_{\mu}-i\left[A_{\mu},A_{\nu}\right]$ and
$D_{\mu}=\partial_{\mu}-iA_{\mu}$.
The masses of the gauge boson and the Higgs boson are given by,
\begin{eqnarray}
&&m_{_W}=\sqrt{\frac{g^2}{2\lambda}} v,\quad m_{_H}=\frac{1}{\sqrt{2}} v.
\end{eqnarray}
This theory is known not to have any finite-size instanton
solutions, in spite of its importance.
In this section, we will construct instanton-like configurations
for this theory, relevant for tunneling phenomena, including
the baryon and lepton number violation processes.

\subsection{Valley instanton}
The valley equation for this system is given by,
\begin{eqnarray}
&&\frac{\delta^2 S}{\delta A_{\mu}\delta A_{\nu}}F^A_{\nu}
+\frac{\delta^2 S}{\delta A_{\mu}\delta H^{\dagger}}F^H
+\frac{\delta^2 S}{\delta A_{\mu}\delta H}F^{H\dagger}
=\lambda_e F^A_{\mu}\nonumber,
\\
&&\frac{\delta^2 S}{\delta H^{\dagger}\delta A_{\mu}}
F^A_{\mu}+
\frac{\delta^2 S}{\delta H^{\dagger}\delta H}F^{H\dagger}
+\frac{\delta^2 S}{\delta H^{\dagger}\delta H^{\dagger}}F^H
=\lambda_e F^{H\dagger},
\label{valley}
\\
&&F^A_{\mu}=\frac{\delta S}{\delta A_{\mu}},
\quad F^H =\frac{\delta S}{\delta H},\nonumber
\end{eqnarray}
where the integration over the space-time is implicit.
The valley is parametrized by the eigenvalue $\lambda_e$ that
is identified with the zero mode corresponding to the scale invariance
in the massless limit, $v\rightarrow 0$.

To simplify the equation, we adopt the following ansatz;
\begin{eqnarray}
A_{\mu}(x)=\frac{x_{\nu}\bar{\sigma}_{\mu\nu}}{x^2}\cdot 2a(r),
\quad H(x)=v\left( 1-h(r)\right)\eta,
\label{ansatz}
\end{eqnarray}
where $\eta$ is a constant isospinor, and $a$ and
$h$ are real dimensionless functions of dimensionless variable
$r$, which is defined by $r=\sqrt{x^2}/\rho$. The matrix
$\bar{\sigma}_{\mu\nu}$ is defined, according to the conventions of
Ref.\cite{Esp}, as $\bar{\sigma}_{\mu\nu}=\bar{\eta}_{a\mu\nu}\sigma^a/2$.
We have introduced the scaling parameter $\rho$ so that we adjust the
radius of valley instanton as we will see later in
subsection 4.3. The tensor structure
in (\ref{ansatz}) is the same as that of the instanton  in the singular
gauge \cite{thooft}.

Inserting this ansatz to (\ref{valley}), the structure
of $F^A_{\mu}$ and $F^{H\dagger}$ is
determined as the following;
\begin{eqnarray}
F^A_{\mu}(x)=\frac{x_{\nu}\bar{\sigma}_{\mu\nu}}{x^2}\cdot
\frac{2v^2}{\lambda}f^a(r),
&&F^{H\dagger}(x)=-\frac{v^3}{\lambda}f^h(r)\eta.
\label{fansatz}
\end{eqnarray}
By using this ansatz, (\ref{ansatz}) and (\ref{fansatz}), the
valley equation (\ref{valley}) is reduced to the following;
\begin{eqnarray}
&&-\frac{1}{r}\frac{d}{dr}\left(r\frac{da}{dr}\right)
+\frac{4}{r^2}a(a-1)(2a-1)
+\frac{g^2}{2\lambda}\rhovev a(1-h)^2=\mr(\rho
v)^2 f^a,
\label{eqn:va}\\
&&-\frac{1}{r^3}\frac{d}{dr}\left(r^3\frac{dh}{dr}\right)
+\frac{3}{r^2}(h-1)a^2
+\frac{1}{4}\rhovev h(h-1)(h-2)=\rhovev f^h,
\label{eqn:vh}\\
&&-\frac{1}{r}\frac{d}{dr}\left(r\frac{df^a}{dr}\right)
+\frac{4}{r^2}(6a^2-6a+1)f^a+\frac{g^2}{2\lambda}\rhovev (h-1)^2f^a
\nonumber
\\
&&\hspace{30ex}+\mr\rhovev a(h-1)f^h=\mr\rhovev \nu f^a,
\label{eqn:vfa}\\
&&-\frac{1}{r^3}\frac{d}{dr}\left(r^3\frac{df^h}{dr}\right)
+\frac{3a^2}{r^2}f^h+\frac{1}{4}\rhovev (3h^2-6h+2)f^h
\nonumber
\\
&&\hspace{30ex}+\frac{6a}{r^2}(h-1)f^a=\rhovev \nu f^h,
\label{eqn:vfh}
\end{eqnarray}
where $\nu$ is defined as $\lambda_e=v^2\nu/\lambda$.

In the massless limit, $\rho v \rightarrow 0$, (\ref{eqn:va}) and
(\ref{eqn:vh}) reduce to the equation of motion and (\ref{eqn:vfa})
and (\ref{eqn:vfh}) to the equation for the zero-mode fluctuation around the
instanton solution. The solution of this set of equations is the following;
\begin{eqnarray}
\begin{array}{cc}
\displaystyle a_{0}=\frac{1}{1+r^{2}},
&\displaystyle h_{0}=1-\left(\frac{r^{2}}{1+r^{2}}\right)^{1/2},
\\[4mm]
\displaystyle f^a_{0}=\frac{2Cr^{2}}{(1+r^{2})^{2}},
&\displaystyle f^h_{0}=\frac{Cr}{(1+r^{2})^{3/2}},
\end{array}
\label{eq:a0}
\end{eqnarray}
where $C$ is an arbitrary function of $\rho v$.
Note that $a_{0}$ is an instanton solution in the singular gauge and
$h_{0}$ is a Higgs configuration in the instanton background \cite{thooft}.
We have adjusted the scaling parameter $\rho$ so that the radius of the
instanton solution is unity. The mode solutions
$f^a_{0}$ and $f^h_{0}$ are obtained from $\partial a_0/\partial \rho$ and
$\partial h_0/\partial \rho$, respectively.

Now we will construct the valley instanton analytically.
When $\rho v=0$, it is given by the ordinary instanton configuration $a_{0}$,
$h_{0}$, $f^{a}_{0}$ and $f^{h}_{0}$.
When $\rho v$ is small but not zero, it is expected that small $\rho v$
corrections appear in the solution.
On the other hand, at large distance from the core of the valley
instanton, this solution is expected to decay exponentially, because
gauge boson and Higgs boson are massive.
Therefore, the solution is similar to the
instanton near the origin and decays exponentially in the asymptotic region.
In the following, we will solve the valley equation in both regions
and analyze the connection in the intermediate region.
In this manner we will find the solution.

In the asymptotic region, $a$, $h$, $f^{a}$
and $f^{h}$ become small and the valley equation can be linearized;
\begin{eqnarray}
&&-\frac{1}{r}\frac{d}{dr}\left(r\frac{da}{dr}\right)
+\frac{4}{r^{2}}a+\frac{g^{2}}{2\lambda}(\rho v)^{2}a
=\frac{g^{2}}{\lambda}(\rho v)^{2}f^{a},
\label{eqn:la}\\
&&-\frac{1}{r^{3}}\frac{d}{dr}\left(r^{3}\frac{dh}{dr}\right)
+\frac{1}{2}(\rho v)^{2}h=(\rho v)^{2}f^{h},
\label{eqn:lh}\\
&&-\frac{1}{r}\frac{d}{dr}\left(r\frac{df^{a}}{dr}\right)
+\frac{4}{r^{2}}f^{a}+\frac{g^{2}}{2\lambda}(\rho v)^{2}f^{a}
=\frac{g^{2}}{\lambda}(\rho v)^{2}\nu f^{h},
\label{eqn:lfa}\\
&&-\frac{1}{r^{3}}\frac{d}{dr}\left(r^{3}\frac{df^{h}}{dr}\right)
+\frac{1}{2}(\rho v)^{2}f^{h}=(\rho v)^{2}\nu f^{h}.
\label{eqn:lfh}
\end{eqnarray}
The solution of this set of equations is
\begin{eqnarray}
&&a(r)=C_{1}\,r\frac{d}{dr}G_{\rho m_{_W}}(r)+\frac{1}{\nu}f^{a}(r),\\
&&h(r)=C_{2}\,G_{\rho m_{_H}}(r)+\frac{1}{\nu}f^{h}(r),\\
&&f^{a}(r)=C_{3}\,r\frac{d}{dr}G_{\rho \mu_{_W}}(r),\\
&&f^{h}(r)=C_{4}\,G_{\rho \mu_{_H}}(r),
\end{eqnarray}
where $C_{i}$ are arbitrary functions of $\rho v$ and $\mu_{_{W,\,H}}$
are defined as
$\mu_{_{W,\,H}}=m_{_{W,\,H}}\sqrt{1-2\nu}$.
As was expected above, these solutions decay exponentially at
infinity and
when $r\ll(\rho
v)^{-1}$ they have the series expansions;
\begin{eqnarray}
&&a(r)=\frac{C_{1}}{(2\pi)^{2}}
\left[-\frac{2}{r^{2}}+\frac{1}{2}(\rho m_{_W})^{2}+\cdots\right]
+\frac{C_{3}}{\nu(2\pi)^{2}}
\left[-\frac{2}{r^{2}}+\frac{1}{2}(\rho \mu_{_W})^{2}+\cdots\right],
\label{eqn:aa}\\
&&h(r)=\frac{C_{2}}{(2\pi)^{2}}
\left[\frac{1}{r^{2}}+\frac{1}{2}(\rho m_{_H})^{2}\ln(\rho m_{_H}rc)
+\cdots\right]\nonumber\\
&&\hspace{30ex}+\frac{C_{4}}{\nu(2\pi)^{2}}
\left[\frac{1}{r^{2}}+\frac{1}{2}(\rho \mu_{_H})^{2}\ln(\rho \mu_{_H}rc)
+\cdots\right],
\label{eqn:ah}\\
&&f^{a}(r)=\frac{C_{3}}{(2\pi)^{2}}
\left[-\frac{2}{r^{2}}+\frac{1}{2}(\rho \mu_{_W})^{2}+\cdots\right],
\label{eqn:afa}\\
&&f^{h}(r)=\frac{C_{4}}{(2\pi)^{2}}
\left[\frac{1}{r^{2}}+\frac{1}{2}(\rho \mu_{_H})^{2}\ln(\rho \mu_{_H}rc)
+\cdots\right],
\label{eqn:afh}
\end{eqnarray}
$c$ being a numerical constant $e^{\gamma-1/2}/2$, where $\gamma$ is
the Euler's constant.

Near the origin, we expect that the valley instanton is similar to the
ordinary instanton.
Then the following replacement of the field variables is convenient;
$a=a_{0}+(\rho v)^{2}\hat{a}$,
$h=h_{0}+(\rho v)^{2}\hat{h}$,
$f^{a}=f^{a}_{0}+(\rho v)^{2}\hat{f^{a}}$,
$f^{h}=f^{h}_{0}+(\rho v)^{2}\hat{f^{h}}$.
If we assume
$a_{0}\gg(\rho v)^{2}\hat{a}$,
$h_{0}\gg(\rho v)^{2}\hat{h}$,
$f^{a}_{0}\gg(\rho v)^{2}\hat{f^{a}}$
and $f^{h}_{0}\gg(\rho v)^{2}\hat{f^{h}}$,
the valley equation becomes
\begin{eqnarray}
&&-\frac{1}{r}\frac{d}{dr}\left(r\frac{d\hat{a}}{dr}\right)
+\frac{4}{r^{2}}(6a_{0}^{2}-6a_{0}+1)\hat{a}
+\frac{g^{2}}{2\lambda}a_{0}(h_{0}-1)^{2}
=\frac{g^{2}}{\lambda}f^{a}_{0},
\label{eqn:ha}\\
&&-\frac{1}{r^{3}}\frac{d}{dr}\left(r^{3}\frac{d\hat{h}}{dr}\right)
+\frac{3}{r^{2}}a_{0}^{2}\hat{h}+\frac{6}{r^{2}}(h_{0}-1)a_{0}\hat{a}
+\frac{1}{4} h_{0}(h_{0}-1)(h_{0}-2)=f^{h}_{0},
\label{eqn:hh}\\
&&-\frac{1}{r}\frac{d}{dr}\left(r\frac{d\hat{f^{a}}}{dr}\right)
+\frac{4}{r^{2}}(6a_{0}^{2}-6a_{0}+1)\hat{f^{a}}
+\frac{24}{r^{2}}(2a_{0}-1)f^{a}_{0}\hat{a}\nonumber\\
&&\hspace{28ex}+\frac{g^{2}}{2\lambda}(h_{0}-1)^{2}f^{a}_{0}
+\frac{g^{2}}{\lambda}a_{0}(h_{0}-1)f^{h}_{0}
=\frac{g^{2}}{\lambda}\nu f^{a}_{0},
\label{eqn:hfa}\\
&&-\frac{1}{r^{3}}\frac{d}{dr}\left(r^{3}\frac{d\hat{f^{h}}}{dr}\right)
+\frac{3}{r^{2}}a_{0}^{2}\hat{f^{h}}
+\frac{6}{r^{2}}a_{0}f^{h}_{0}\hat{a}
+\frac{1}{4}(3h_{0}^{2}-6h_{0}+2)f^{h}_{0}\nonumber
\\
&&\hspace{20ex}+\frac{6}{r^{2}}a_{0}(h_{0}-1)\hat{f^{a}}
+\frac{6}{r^{2}}(h_{0}-1)f^{a}_{0}\hat{a}
+\frac{6}{r^{2}}a_{0}f^{a}_{0}\hat{h}
=\nu f^{h}_{0}.
\label{eqn:hfh}
\end{eqnarray}
To solve this equation, we introduce solutions of the following
equations;
\begin{equation}
\begin{array}{l}
\displaystyle
-\frac{1}{r}\frac{d}{dr}\left(r\frac{d\varphi_{a}}{dr}\right)
+\frac{4}{r^{2}}(6a_{0}^{2}-6a_{0}+1)\varphi_{a}=0,
\\
\\
\displaystyle
-\frac{1}{r^{3}}\frac{d}{dr}\left(r^{3}\frac{d\varphi_{h}}{dr}\right)
+\frac{3}{r^{2}}a_{0}^{2}\varphi_{h}=0.
\label{eqn:zh}
\end{array}
\end{equation}
They are given as,
\begin{equation}
\varphi_{a}=\frac{r^{2}}{(1+r^{2})^{2}},
\quad\varphi_{h}=\left(\frac{r^{2}}{1+r^{2}}\right)^{1/2}.
\end{equation}
Using these solutions, we will integrate the valley equation.
We multiply (\ref{eqn:ha}) and (\ref{eqn:hfa}) by $r\varphi_{a}$, and
multiply (\ref{eqn:hh}) and (\ref{eqn:hfh}) by $r^{3}\varphi_{h}$ then
integrate them from $0$ to $r$.
Integrating by parts and using (\ref{eqn:zh}), we obtain
\begin{eqnarray}
&&-\varphi_{a}r\frac{d\hat{a}}{dr}+\frac{d\varphi_{a}}{dr}r\hat{a}
=\frac{g^{2}}{\lambda}\int_{0}^{r}dr' r'\varphi_{a}
\left[f^{a}_{0}-\frac{1}{2}a_{0}(h_{0}-1)^{2}\right],
\label{eqn:dha}\\
&&-\varphi_{h}r^{3}\frac{d\hat{h}}{dr}+\hat{h}r^{3}\frac{d\varphi_{h}}{dr}
=\int_{0}^{r}dr' r'^{3}\varphi_{h}
\left[f^{h}_{0}-\frac{1}{4}h_{0}(h_{0}-1)(h_{0}-2)
-\frac{6}{r'^{2}}(h_{0}-1)a_{0}\hat{a}\right],
\nonumber\\
\label{eqn:dhh}\\
&&-\varphi_{a}r\frac{d\hat{f^{a}}}{dr}
+\frac{d\varphi_{a}}{dr}r\hat{f^{a}}\nonumber\\
&&\hspace{2ex}=\int_{0}^{r}dr' r'\varphi_{a}
\left[\frac{g^{2}}{\lambda}\nu f^{a}_{0}
-\frac{24}{r'^{2}}(2a_{0}-1)f^{a}_{0}\hat{a}
-\frac{g^{2}}{2\lambda}(h_{0}-1)^{2}f^{a}_{0}
-\frac{g^{2}}{\lambda}a_{0}(h_{0}-1)f^{h}_{0}\right],
\label{eqn:dhfa}\\
&&-\varphi_{h}r^{3}\frac{d\hat{f^{h}}}{dr}
+\hat{h}r^{3}\frac{d\varphi_{h}}{dr}\nonumber\\
&&\hspace{2ex}=\int_{0}^{r}dr' r'^{3}\varphi_{h}
\left[\nu f^{h}_{0}-\frac{6}{r'^{2}}a_{0}f^{a}_{0}\hat{a}
-\frac{1}{4}(3h_{0}^{2}-6h_{0}+2)f^{h}_{0}\right.\nonumber\\
&&\hspace{29ex}\left.-\frac{6}{r'^{2}}a_{0}(h_{0}-1)\hat{f^{a}}
-\frac{6}{r'^{2}}(h_{0}-1)f^{a}_{0}\hat{a}
-\frac{6}{r'^{2}}a_{0}f^{a}_{0}\hat{h}\right].
\label{eqn:dhfh}
\end{eqnarray}

First we will find $\hat{a}$.
The right-hand side of (\ref{eqn:dha}) is proportional to $(C-1/4)$
and when $r$ goes to infinity this approaches a constant.
At $r\gg1$, (\ref{eqn:dha}) becomes
\begin{equation}
-\frac{1}{r}\frac{d\hat{a}}{dr}-\frac{2}{r^{2}}\hat{a}
=\frac{1}{3}\frac{g^{2}}{\lambda}\left(C-\frac{1}{4}\right).
\end{equation}
Then at $r\gg1$, $\hat{a}(r)$ is proportional to
$(C-1/4)r^{2}$ and $a(r)$ becomes
\begin{equation}
a=\frac{1}{r^{2}}
-\frac{(\rho v)^{2}g^{2}}{12\lambda}\left(C-\frac{1}{4}\right)r^{2}+\cdots.
\end{equation}
To match this with (\ref{eqn:aa}), it must be hold that $C=1/4$ when
$\rho v=0$.
When $C=1/4$, the right-hand sides of (\ref{eqn:dha}) vanishes and
$\hat{a}$ satisfy $-\varphi_{a}d\hat{a}/dr+\hat{a}d\varphi_{a}/dr=0$.
Hence $\hat{a}$ is $\hat{a}=D\,\varphi_{a}$, where $D$ is a constant.
Identifying $a_{0}+(\rho v)^{2}\hat{a}$ with (\ref{eqn:aa}) again at $r\gg1$,
we find that
$C_{1}+C_{3}/\nu=-2\pi^{2}$ and $C_{3}=-\pi^{2}$ at $\rho v=0$.
In the same manner, $\hat{h}$,
$\hat{f^{a}}$ and $\hat{f^{h}}$ are obtained.
At $r\gg1$, we find
\begin{eqnarray}
&&\hat{h}={\rm const.}+\cdots,
\nonumber\\
&&\hat{f^{a}}=\frac{g^{2}}{48\lambda}\left(\frac{1}{4}-\nu\right)r^{2}
-\frac{g^{2}}{16\lambda}(1-2\nu)+\cdots,
\\
&&\hat{f^{h}}=\frac{1}{16}(1-2\nu)\ln r+\cdots.
\nonumber
\end{eqnarray}
Here ${\rm const.}$ is a constant of integration.
Comparing (\ref{eqn:ah})-(\ref{eqn:afh}) with
them, we find that $C_{2}+C_{4}/\nu=2\pi^{2}$,
$C_{4}=\pi^{2}$ and $\nu=1/4$ at $\rho v=0$.

Now we have obtained the solution of the new valley equation. Near the
origin of the valley instanton,
$r \ll (\rho m_{_{W,H}})^{-1/2}$,
it is given by,
\begin{eqnarray}
\begin{array}{ll}
\displaystyle a(r)=\frac{1}{1+r^{2}},
&\displaystyle \hspace{8ex}h(r)=1-\left(\frac{r^{2}}{1+r^{2}}\right)^{1/2},
\\[4mm]
\displaystyle f^a(r)=\frac{r^{2}}{2(1+r^{2})^{2}},
&\displaystyle \hspace{8ex}f^h(r)=\frac{r}{4(1+r^{2})^{3/2}},
\end{array}
\label{eq:vnear}
\end{eqnarray}
where we ignore the correction terms that go to zero as $\rho v
\rightarrow 0$, since they are too small comparing with the leading terms.
As $r$ becomes larger, the leading terms are getting smaller
and so the correction terms become more important;
\begin{equation}
\begin{array}{ll}
\displaystyle a(r)=\frac{1}{r^{2}}+o((\rho v)^2),
&\displaystyle h(r)=\frac{1}{2r^{2}}-\frac{(\rho v)^{2}}{16}\ln2+\cdots,
\\[4mm]
\displaystyle f^{a}(r)=\frac{1}{2r^{2}}
-\frac{g^{2}(\rho v)^{2}}{32\lambda}+\cdots,
&\displaystyle f^{h}(r)=\frac{1}{4r^{2}}-\frac{(\rho v)^{2}}{32}\ln r+\cdots,
\end{array}
\label{eq:vcorr}
\end{equation}
for $(\rho m_{_{W,H}})^{-1/2} \ll r \ll (\rho m_{_{W,H}})^{-1}$.
Finally, far from the origin, $r \gg (\rho m_{_{W,H}})^{-1/2}$,
the solution is given by the following:
\begin{equation}
\begin{array}{l}
\displaystyle a(r)=2\pi^2\, r\frac{d}{dr}G_{\rho m_{_W}}(r)
+\frac{1}{\nu}f^{a}(r),\\
\displaystyle h(r)=-2\pi^2\,G_{\rho m_{_H}}(r)+\frac{1}{\nu}f^{h}(r),\\
\displaystyle f^{a}(r)=-\pi^2\,r\frac{d}{dr}G_{\rho \mu_{_W}}(r),\\
\displaystyle f^{h}(r)=\pi^2\,G_{\rho \mu_{_H}}(r),\\
\end{array}
\label{eq:vfar}
\end{equation}
where $\nu = 1/4$ for $\rho v = 0$. In (\ref{eq:vfar}), we ignore
correction terms, since they are too small.

Let us make a brief comment about the consistency of our analysis.
Until now, we have implicitly assumed that there exists an overlapping
region where both (\ref{eqn:la})-(\ref{eqn:lfh}) and
(\ref{eqn:ha})-(\ref{eqn:hfh}) are valid.
Using the above solution, it is found that (\ref{eqn:la})-(\ref{eqn:lfh})
are valid when $r\gg (\rho m_{_{W,H}})^{-1/2}$ and
(\ref{eqn:ha})-(\ref{eqn:hfh})
are valid when $r\ll (\rho m_{_{W,H}})^{-1}$.
If $\rho m_{_{W,H}}$ is small enough,
there exists the overlapping region $(\rho m_{_{W,H}})^{-1/2}\ll r\ll (\rho
m_{_{W,H}})^{-1}$.
Then our analysis is consistent.

The action of the valley instanton can be calculated using the above
solution.
Rewriting the action in terms of  $a$ and $h$, we find
\begin{eqnarray}
&&S_{g}=\frac{12\pi^{2}}{g^{2}}
\int_{0}^{\infty}\frac{dr}{r}\left\{
\left( r\frac{da}{dr}\right)^{2}+4a^{2}(a-1)^{2}
\right\},
\\
&&S_{h}=\frac{2\pi^{2}}{\lambda}(\rho v)^{2}\int_{0}^{\infty}
r^{3}dr\left\{
\left(\frac{dh}{dr}\right)^{3}+\frac{3}{r^{2}}(h-1)^{2}a^{2}
+\frac{1}{8}(\rho v)^{2}h^{2}(h-2)^{2}\right\}.
\end{eqnarray}
Substituting the above solution for $S$, we obtain
\begin{equation}
\label{eqn:aaction}
S=\frac{8\pi^{2}}{g^{2}}
    +\frac{2\pi^{2}}{\lambda}(\rho v)^{2}-\frac{\pi^2}{4\lambda}(\rho
v)^{4}\ln(\rho v)+O((\rho v)^4).
\end{equation}
The leading contribution $8\pi^{2}/g^{2}$ comes from $S_{g}$ for $a_{0}$,
which is the action of the instanton, and the next-to-leading and the third
contributions come from $S_{h}$ for $a_{0}$ and $h_{0}$.

\subsection{Constrained instanton}
In this subsection, we consider the constrained instanton.
According to Affleck's analysis \cite{Aff}, the constrained instanton satisfies
the following equation:
\begin{eqnarray}
&&\frac{\delta S}{\delta A_{\mu}}+\sigma\frac{\delta O_A}{\delta A_{\mu}}=0,
\\
&&\frac{\delta S}{\delta H}+\sigma\frac{\delta O_H}{\delta H}=0,
\end{eqnarray}
where $\sigma$ is a Lagrange multiplier and depends on the constraint.
Both $O_A$ and $O_H$ are functionals of $A_{\mu}$ and $H$
respectively
that give a solution of the constrained equation, as the scalar theory
in the subsection 3.2.
Here we adopt the ansatz (\ref{ansatz}) again. By the similar analysis
as the valley instanton, it turns out that the behavior of the
constrained instanton is the following. Near the origin of the
instanton, the solution is given by $a_0$, $h_0$ in (\ref{eq:a0}) as
well as the valley instanton. In the region
where $(\rho m_{_{W,H}})^{-1/2} \ll r \ll (\rho m_{_{W,H}})^{-1}$, the solution
is given
by
\begin{equation}
a(r)=\frac{1}{r^{2}}-\frac{\rhovev}{8}\mr+\cdots,
\hspace{5ex}h(r)=\frac{1}{2r^{2}}+\frac{(\rho v)^{2}}{8}\ln(\rho v rc)+\cdots.
\label{eq:ccorr}
\end{equation}
Let us compare (\ref{eq:vcorr}) and (\ref{eq:ccorr}). The correction
term of the valley instanton is smaller than one of the constrained instanton.
Finally, for $r \gg (\rho m_{_{W,H}})^{-1/2} $, the constrained instanton is
given by
\begin{eqnarray}
a(r)=-2 \pi^2 r\frac{d}{dr}G_{\rho m_{_W}}(r),
\hspace{5ex}h(r)=2 \pi^2 G_{\rho m_{_H}}(r).
\end{eqnarray}
The action of the constrained instanton is the following:
\begin{equation}
S=\frac{8\pi^{2}}{g^{2}}
    +\frac{2\pi^{2}}{\lambda}(\rho v)^{2}+O((\rho v)^4 \ln(\rho v)).
\label{eq:conaction}
\end{equation}
The leading contribution $8\pi^{2}/g^{2}$ comes from $S_{g}$ for $a_{0}$,
 which is the action of the instanton, and the next-to-leading
contributions comes from $S_{h}$ for $a_{0}$ and $h_{0}$. The
difference from the valley instanton is that we cannot determine the
term of $O((\rho v)^4\ln(\rho v))$ by the current analysis.

Now we choose a constraint and analyze the constrained instanton
. We adopt the following functionals for the constraint:
$O_A=i g^2\int d^4 x {\rm tr}F_{\mu\nu}F_{\nu\rho}F_{\rho\mu},\
O_H=0$. This constraint is one of the simplest for giving the constrained
instanton.
Then the constrained equation of motion is given by,
\begin{eqnarray}
&&-\frac{1}{r}\frac{d}{dr}\left(r\frac{da}{dr}\right)
+\frac{4}{r^2}a(a-1)(2a-1)
+\frac{g^2}{2\lambda}\rhovev a(1-h)^2\nonumber\\
&&~\hspace{5ex}+6\rhovev
\frac{\tilde{\sigma}}{r^2}\Biggl\{
(2a-1)\left(\frac{da}{dr}\right)^2
+2a(a-1)\frac{d^2 a}{dr^2}
 -\frac{2}{r}a(a-1)\frac{da}{dr}\label{eqn:ca}\\
&&~\hspace{20ex} -\frac{4}{r^2}a^2(a-1)^2(2a-1)\Biggr\}=0
,\nonumber\\
&&-\frac{1}{r^3}\frac{d}{dr}\left(r^3\frac{dh}{dr}\right)
+\frac{3}{r^2}(h-1)a^2
+\frac{1}{4}\rhovev h(h-1)(h-2)=0,
\label{eqn:ch}
\end{eqnarray}
where $\sigma=\rho^4v^2\tilde{\sigma}$.
By solving these equations approximately, in fact, we obtain the
behavior of the solution that we have given previously, and  the
multiplier is determined by $\tilde{\sigma}=5g^2/48\lambda $ as $\rho
v \rightarrow 0$.

\subsection{Numerical analysis}
In this subsection, we solve the valley equation (\ref{eqn:va})-(\ref{eqn:vfh})
and the constrained equation (\ref{eqn:ca}), (\ref{eqn:ch})
numerically, and compare the valley instanton and constrained
instanton.

We need a careful discussion for solving the valley equation
(\ref{eqn:va})-(\ref{eqn:vfh}):
Since the solution must be regular at the origin, we assume the following
expansions for $r \ll 1$;
\begin{eqnarray}
\begin{array}{cc}
\displaystyle a(r)=\sum_{n=0}^{\infty} a_{(n)} r^n,
&\displaystyle h(r)=\sum_{n=0}^{\infty} h_{(n)} r^n,
\\
\displaystyle f^a(r)=\sum_{n=0}^{\infty} f^a_{(n)} r^n,
&\displaystyle f^h(r)=\sum_{n=0}^{\infty} f^h_{(n)} r^n.
\end{array}
\label{expand}
\end{eqnarray}
Inserting (\ref{expand}) to (\ref{eqn:va})-(\ref{eqn:vfh}), we obtain
\begin{eqnarray}
&&a_{(0)}=1,h_{(0)}=1,f^a_{(0)}=0,f^h_{(0)}=0,
\nonumber
\\
&&a_{(1)}=0,f^a_{(1)}=0,
\\
&&h_{(2)}=0,f^h_{(2)}=0.
\nonumber
\end{eqnarray}
The coefficients $a_{(2)}$, $h_{(1)}$, $f^a_{(2)}$ and
$f^h_{(1)}$ are not determined and remain as free parameters.
The higher-order coefficients ($n \ge 3$) are determined
in terms of these parameters.
Four free parameters are determined by boundary conditions at infinity.
The finiteness of action requires
$a$, $h$ $\rightarrow 0$ faster than $1/r^2$ at infinity. This
condition also requires $f_a$, $f_h$ $\rightarrow 0$.

We have introduced $\rho$ as a free scale parameter.
We adjust this parameter $\rho$ so that $a_{(2)}=-2$
to make the radius of the valley instanton unity.
As a result we have four parameters $h_{(1)}$, $f^a_{(2)}$,
$f^h_{(1)}$ and $\rho v$ for a given $\nu$.
These four parameters are determined so that $a$, $h$, $f^a$,
and $f^h$ $\rightarrow 0$ at infinity. In the case
of the constrained instanton, the two parameters $h_{(1)}$ and
$\tilde{\sigma}$ are determined under $a_{(2)}=-2$ so that $a$, $h$
$\rightarrow 0$ at infinity.

\begin{wrapfigure}{c}{14cm}
\centerline{
\epsfxsize=10cm
\epsfbox{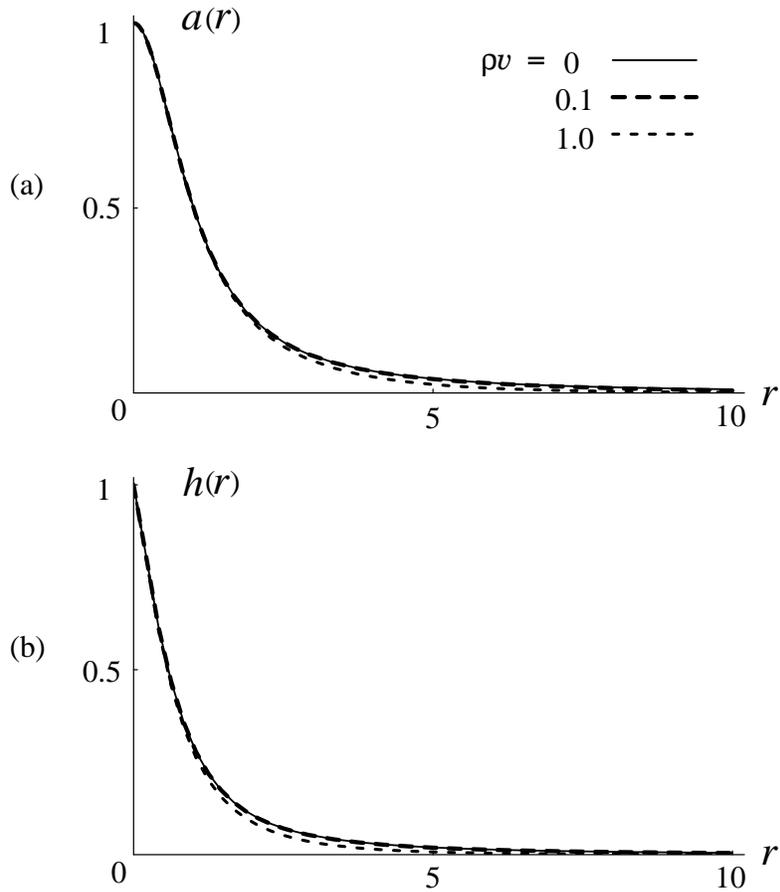}
}
\caption{
Shapes of the numerical solution of the valley equation, $a(r)$ and
$h(r)$ for $\rho v =0.1, 1.0$ near the origin.
The solid lines denote the original instanton solution, $a_0$
and $h_0$.
}
\label{fig:shval}
\end{wrapfigure}

A numerical solution of the valley equation near the origin is plotted
in Fig.\ref{fig:shval} for $\rho v=0.1,\ 1.0$ at $\lambda /g^2 =1$,
when $m_{_W}=m_{_H}$.
We plot the instanton solution (\ref{eq:a0}) by the solid line,
which corresponds to the valley instanton for $\rho v=0$.
This behavior of the numerical solution for $\rho v=0.1$
agrees with the result of the previous subsection, (\ref{eq:vnear}).
Moreover, even when
$\rho v=1.0$, the numerical solution is quite similar to the instanton
solution. We find that the behavior of $f^a(r)$ and $f^h(r)$
also agrees with the analytical result (\ref{eq:vnear})
as well as $a(r)$ and $h(r)$.
We also find that the numerical solution
in the asymptotic region where $r \gg (\rho v)^{-1/2}$, is damping
exponentially and agrees with the analytical result (\ref{eq:vfar}).

\eject{}~\vfil
\vfil

\begin{wrapfigure}{c}{14cm}
\centerline{
\epsfxsize=10cm
\epsfbox{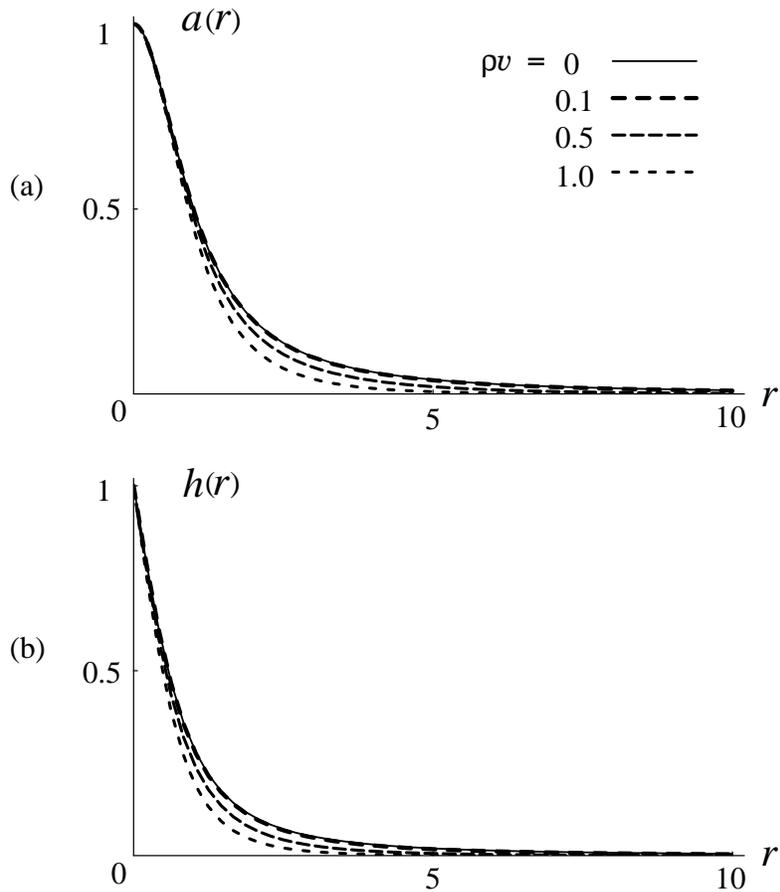}
}
\caption{
Shapes of the numerical solution of the constrained
instanton, $a(r)$ and $h(r)$ for $\rho v =0.1, 0.5, 1.0$ near the
origin. The solid lines denote the original instanton solution, $a_0$
and $h_0$.}
\label{fig:shcon}
\end{wrapfigure}

On the other hand, a solution of the constrained equation is
plotted for $\rho v=0.1,\ 0.5,\ 1.0$ at $\lambda /g^2 =1$
in Fig.\ref{fig:shcon}.
The numerical solution for $\rho v=0.1$ also
agrees with the analytical result. As $\rho v $ is larger,
both the valley instanton and the constrained instanton
are more deformed from the original instanton solution.
Nevertheless, the
correction of the constrained instanton from the original instanton
is much larger than one of the valley instanton, especially
when $\rho v \lsim 0.5$.

\eject
{}~\vfill
\vfill

\begin{wrapfigure}{c}{14cm}
\centerline{
\epsfxsize=12cm
\epsfbox{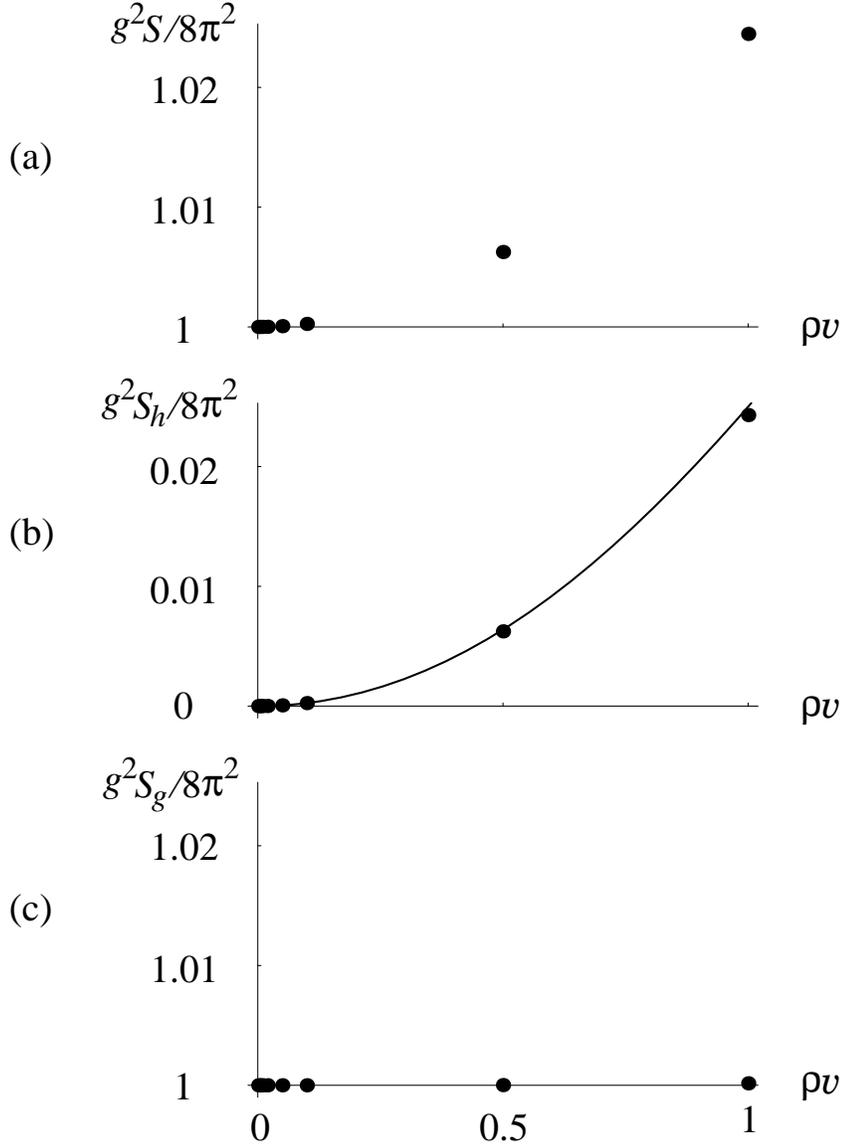}
}
\caption{
(a) The action $S$ (in unit of $g^2 /8 \pi^2$) of the numerical solution
of the valley equation, at $\lambda /g^2 =1$, as a function of the
parameter $\rho v$.
(b) The contribution from the Higgs sector, $S_h$.
The solid line shows the behavior of the analytical result
that $g^2 S_h/8 \pi^2 = (\rho v)^2/4 -(\rho v)^4 \ln (\rho v)/32$.
(c) The contribution from the gauge sector, $S_g$.
}
\label{fig:actval}
\end{wrapfigure}

The values of the action of the valley instanton for $\rho v=0.001\sim
1.0$ are plotted in Fig.\ref{fig:actval}:
Fig.\ref{fig:actval}(a) depicts the behavior of the
total action $S$, while the contribution from the gauge part $S_g$
and from the Higgs part $S_h$ are in (b) and (c) respectively.
The solid line shows the behavior of $S_h$
of the analytical result (\ref{eqn:aaction}). From Fig.\ref{fig:actval}
(c), it turns that $S_g$ is almost independent of $\rho v$.
This figure shows that our numerical solutions are quite consistent
 with (\ref{eqn:aaction}), even when $\rho v = 1.0$.

\eject

\begin{wrapfigure}{c}{14cm}
\centerline{
\epsfxsize=12cm
\epsfbox{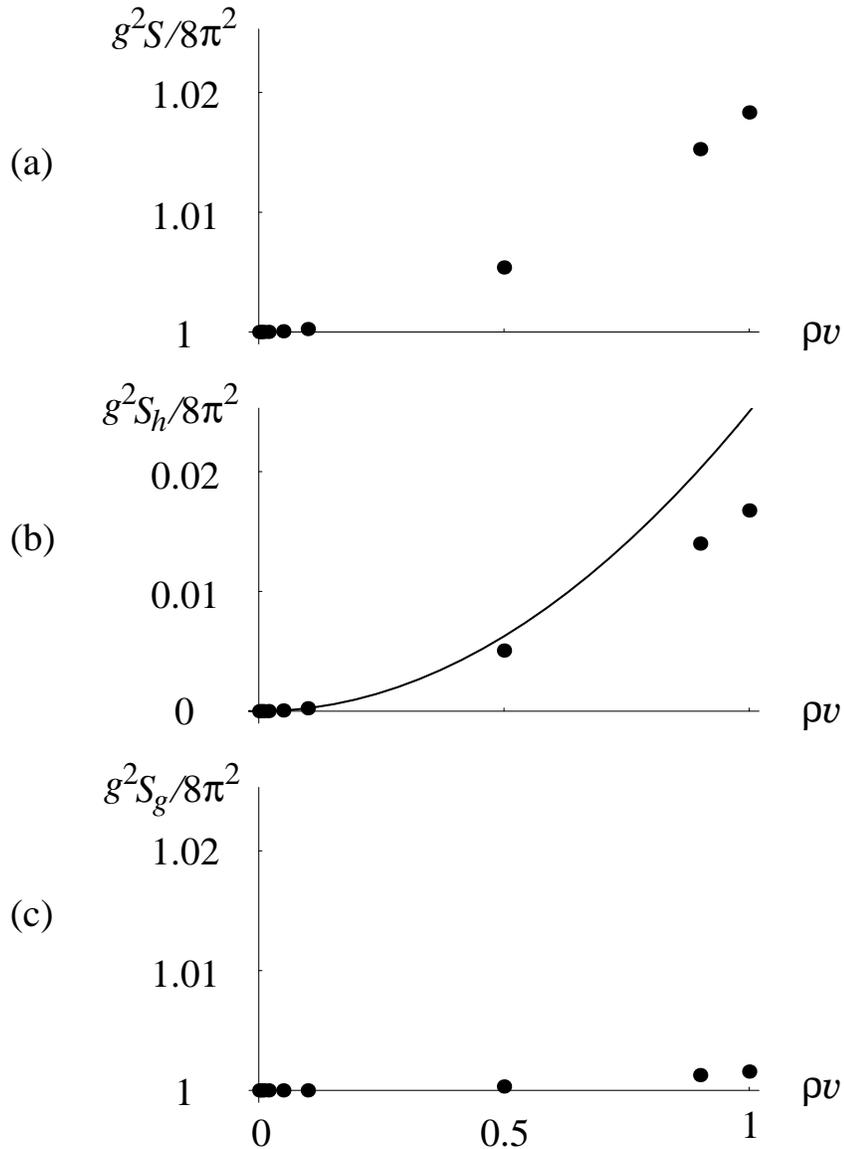}
}
\caption{(a) The action $S$ (in units $g^2 /8 \pi^2$) of the numerical solution
of the constrained instanton, at $\lambda /g^2 =1$, as a function of the
parameter $\rho v$.
(b) The contribution from the Higgs sector, $S_h$.
The solid line shows the behavior of the analytical result
that $g^2 S_h/8 \pi^2 = (\rho v)^2/4$.
(c) The contribution from the gauge sector, $S_g$.
}
\label{fig:actcon}
\end{wrapfigure}

The values of the action of the constrained instanton for $\rho
v=0.001\sim 1.0$ are
plotted in Fig.\ref{fig:actcon}.
When $\rho v \lsim 0.5$, we can no
longer use the analytical result (\ref{eq:conaction}). This is
attributed
to the large deformation from the original instanton. We notice that
the action of the constrained instanton is smaller than that of the
valley instanton for the same value of $\rho v$. This result is
natural at the point that the parameter $\rho v$ corresponds to the
scale parameter.
This point was elaborated upon in section 2, using
the toymodel (\ref{eq:toyconst}).
To repeat, the fact that the constrained instanton
has the smaller action than the valley instanton for the
same value of $\rho v$ does not mean that the constrained
instanton gives more important contribution to the functional integral
than the valley instanton.

We summarize all the numerical data in Table \ref{tab:valresult}
and  \ref{tab:conresult}, for the valley instanton and for the
constrained instanton, respectively.

\begin{wraptable}{c}{14cm}
\begin{center}
\doublerulesep=0pt
\def\arraystretch{1.3}
\begin{tabular}{@{\vrule width 1pt \quad}c
@{\quad \vrule width 1pt \, }c|c|c|c|c @{\ \vrule width 1pt}}
\noalign{\hrule height 1pt}
{$\rho v$ } & {$\nu$} & $h_{(1)}$ & $f^a_{(2)}$ &
$f^h_{(1)}$ & $g^{2}S/8\pi^2$ \\
\hline
0.001 & 0.2500 & -1.000 & 0.5001 & 0.2500 & 1.000000 \\
\hline
0.005 & 0.2500 & -1.000 & 0.5001 & 0.2500 & 1.000006 \\
\hline
0.01 & 0.2501  & -1.000 & 0.5001 & 0.2500 & 1.000025 \\
\hline
0.02 & 0.2501  & -1.000 & 0.5003 & 0.2501 & 1.000100 \\
\hline
0.05 & 0.2503  & -1.000 & 0.5014 & 0.2505 & 1.000625 \\
\hline
0.1  & 0.2509  & -1.000 & 0.5044 & 0.2515 & 1.002503 \\
\hline
0.5 & 0.2537   & -1.007 & 0.5478 & 0.2669 & 1.062550\\
\hline
1.0 & 0.2521   & -1.021 & 0.6226 & 0.2926 & 1.244646\\
\noalign{\hrule height 1pt}
\end{tabular}
\end{center}
\caption{The numerical data of the valley instanton in the gauge-Higgs
system.}
\label{tab:valresult}
\end{wraptable}

\begin{wraptable}{c}{14cm}
\begin{center}
\doublerulesep=0pt
\def\arraystretch{1.3}
\begin{tabular}{@{\vrule width 1pt \quad}c @{\quad \vrule width 1pt \, }c|c|c|c
@{\ \vrule width 1pt}}
\noalign{\hrule height 1pt}
{$\rho v$ } & {$\tilde{\sigma}$} & $h_{(1)}$ & {$\rho_0 v$} &
 $g^{2}S/8\pi^2$ \\
\hline
0.001 & 0.1042 & -1.000 & 0.000100   & 1.000000\\
\hline
0.005 & 0.1042 & -1.000 & 0.005000   & 1.000006\\
\hline
0.01  & 0.1041 & -1.000 & 0.010000   & 1.000025\\
\hline
0.02  & 0.1040 & -1.000 & 0.019998   & 1.000100 \\
\hline
0.05  & 0.1035 & -1.002 & 0.049972  & 1.000623 \\
\hline
0.1   & 0.1019 & -1.005 & 0.099779  & 1.002471 \\
\hline
0.5   & 0.0798 & -1.070 & 0.481738  & 1.053903 \\
\hline
0.9   & 0.0637 & -1.144 & 0.833000   & 1.152691 \\
\hline
1.0   & 0.0609 & -1.162 & 0.917406   & 1.183336 \\
\noalign{\hrule height 1pt}
\end{tabular}
\end{center}
\caption{The numerical data of the constrained instanton in the
gauge-Higgs system.}
\label{tab:conresult}
\end{wraptable}

In Table \ref{tab:conresult}, we
show the parameter $\rho_0$ used as ``$\rho$'' in Ref.\cite{Aff}.
The parameter $\rho_0$ is defined by,
\begin{eqnarray}
O = c_0 \rho_0^{4-d},
\end{eqnarray}
where $d$ is a dimension of the operator and $c_0$ is a conveniently
chosen constant. For $O_A=i g^2\int d^4 x {\rm
tr}F_{\mu\nu}F_{\nu\rho}F_{\rho\mu}$ and $O_H=0$, we choose
that $c_0=96 \pi^2 /5g^2$ so that $\rho_0 =\rho$ as $\rho v \rightarrow 0$.
We find that $\rho_0$ nearly agrees with $\rho$.

\subsection{Incorporation of fermions}\label{sec:fermion}
Let us introduce fermions to the gauge-Higgs system and analyze the
fermionic zero mode around the valley instanton. The analysis can be
done with the same procedure as the case of the constrained
instanton \cite{Esp}.
According to Ref.\cite{Esp}, we consider the system  with four
left-hand doublets $q_{La}$ and $l_L$,
and seven right-hand singlets $u_{Ra}$, $d_{Ra}$ and
$e_R$, where $a =1,2,3$ is the SU(3) color index.
The fermionic part of the action is given by,
\begin{eqnarray}
&&S_q=\int d^4 x (i q_L^{\dagger}\sigma_{\mu}D_{\mu}q_L
-i u_R^{\dagger}\bar{\sigma}_{\mu}\partial_{\mu} u_R
-i d_R^{\dagger}\bar{\sigma}_{\mu}\partial_{\mu} d_R\\
\nonumber
&& \hspace{10ex}+y_u q_L^{\dagger}\epsilon H^*
u_R-y_d q_L^{\dagger}H d_R +h.c.),\\
&&S_L=\int d^4 x (i l_L^{\dagger}\sigma_{\mu}D_{\mu}l_L
-i e_R^{\dagger}\bar{\sigma}_{\mu}\partial_{\mu} e_R
-y_L l_L^{\dagger} H e_R +h.c.),
\end{eqnarray}
where $\epsilon^{\alpha\beta}=-\epsilon_{\alpha\beta} = i \sigma^2$,
 $\sigma_{\mu}= (\sigma,i)$ and
$\bar{\sigma}_{\mu}= (\sigma,-i)$.
This system can be regarded as the simplified standard model, in which
the U(1) and SU(3) coupling constants are equal to zero and
the Kobayashi-Maskawa matrix is the identity.
The fermionic zero mode around the valley instanton is given by
the following equation;
\begin{eqnarray}
&&i \sigma_{\mu}D_{\mu}q_L+y_u\epsilon H^* u_R -y_d H d_R=0,\\
&&y_u H^T \epsilon q_L - i \bar{\sigma}_{\mu} \partial_{\mu} u_R=0,\\
&&-y_d H^{\dagger} q_L -i \bar{\sigma}_{\mu}\partial_{\mu} d_R=0,
\end{eqnarray}
where $A_{\mu}$ in the covariant derivative $D_{\mu}$ and $H$ are the valley
instanton.
We obtain the zero mode for leptonic sector by replacing $y_d$ and
$y_u$ in the solution of the above equation with $y_L$ and $0$, respectively.
In general, the valley instanton is given by,
\begin{eqnarray}
A_{\mu}(x)=\frac{x_{\nu}U\bar{\sigma}_{\mu\nu}U^{\dagger}}{x^2}\cdot 2a(r),
\quad H(x)=v\left( 1-h(r)\right)\eta,
\end{eqnarray}
where $U$ is a global SU(2) matrix and we choose $\eta_r$ as (0,1).
The functions $a(r)$ and $h(r)$ have been studied in the previous
subsections.
The solution is given by,
\begin{eqnarray}
&&q_{Lr}^{\dot{\alpha}}(x)=-\frac{1}{\pi\rho^3}
x_{\nu}\bar{\sigma}_{\nu}^{\dot{\alpha} \alpha}
U^{\dagger \beta}_{\alpha}\left(-\epsilon_{rs}\eta^{\dagger
s}\eta_{\beta}u_L(r)+\eta_r \epsilon_{\beta
\gamma}\eta^{\dagger\gamma}d_L(r)\right)\\
&&u_{R\alpha}(x)=\frac{i}{2\pi}
\frac{m_u}{\rho}u_R(r)U^{\dagger\beta}_{\alpha}\eta_{\beta}\\
&&d_{R\alpha}(x)=\frac{i}{2
\pi}\frac{m_d}{\rho}d_R(r)U^{\dagger\beta}_{\alpha}\epsilon_{\beta\gamma}
\eta^{\dagger\gamma},
\end{eqnarray}
where the Greek and the roman letters denote indices of spinor and isospinor,
respectively. The masses of fermions are given by $m_{u,d}=y_{u,d}v$.
The functions of $r$, $q_L(r)$, $u_R(r)$ and $d_R(r)$ are
the solution of the following equation;
\begin{eqnarray}
&&\left(4+r\frac{d}{dr}\right)u_L-a(u_L+2 d_L)+\frac{1}{2}(\rho m_u)^2 (1-h)
u_R=0,\\
&&\left(4+r\frac{d}{dr}\right)d_L-a(d_L+2 u_L)+\frac{1}{2}(\rho m_d)^2 (1-h)
d_R=0,\\
&&\frac{1}{r}\frac{d}{dr}u_R+2(1-h)u_L=0,\\
&&\frac{1}{r}\frac{d}{dr}d_R+2(1-h)d_L=0.
\end{eqnarray}
For small $\rho v$, the solution is given  approximately, by the
following;
\begin{eqnarray}
&&u_L(r)=\left\{
    \begin{array}{ll}
\displaystyle \frac{1}{r(r^2+1)^{3/2}},&
\quad \mbox{if} \quad r \ll (\rho m_u)^{-1/2}\ ; \\[0.4cm]
\displaystyle  -2 \pi^2 \frac{1}{r}\frac{d}{dr}G_{\rho m_u}(r),&
\quad \mbox{if} \quad r \gg (\rho m_u)^{-1/2}\ ;
     \end{array} \right.\\
&&u_R(r)=\left\{
    \begin{array}{ll}
\displaystyle \frac{1}{r^2+1},&
\quad \mbox{if} \quad r \ll (\rho m_u)^{-1/2}\ ; \\[0.4cm]
\displaystyle  2 \pi^2 G_{\rho m_u}(r),&
\quad \mbox{if} \quad r \gg (\rho m_u)^{-1/2},
     \end{array} \right.
\end{eqnarray}
and $d_L$ and $d_R$ are obtained by replacing $m_u$ with $m_d$.
If the masses of fermions are too small, the above approximative
behavior is correct even though $\rho v$ is $O(1)$.

\section{Conclusion and Discussion}
In this paper we have examined instanton-like
configurations in the theory where a mass scale prevents the classical
solution with a finite radius.
A natural way to construct the instanton-like configurations is the one
based on the new valley method.
The resulting configuration, which we call ``valley instanton'', has
been shown to have desirable behaviors.
Since a direction along which the action varies most
slowly is chosen as the collective coordinate, it is expected that
this method gives a more plausible approximation than the constrained
instanton.
Indeed this has been assured in the toy model in section \ref{sec:NV}.
In the new valley method, the corresponding quasi-zero mode in
the Gaussian integration is removed completely, then it gives a smooth
extension of the ordinary collective coordinate method in the case
where zero modes exist and this makes the evaluation of the Gaussian
integration more easily.

To clarify the effectiveness of the method in the field theory, detail
comparisons between the valley instanton and the constrained instanton
in the scalar $\phi^4$ theory and the SU(2) gauge-Higgs system have
been carried out.
When the size of the configuration is small,
the valley instanton can be constructed analytically.
The differences from the constrained instanton is rather small in this case.
For the configuration with a large radius, we have
constructed both instantons numerically.
It is found that the remarkable differences between them appear in
this case.
In the scalar $\phi^4$ theory, while the action of the valley
instanton increases monotonously and remains being positive, the one
of the constrained instanton becomes negative when the radius becomes large.
There also exist the differences between them in SU(2)
gauge-Higgs system.

In addition, we found that the valley instanton has a very similar
configuration to the instanton even when the radius is large both in
the scalar $\phi^4$ theory and the SU(2) gauge-Higgs
system.\footnote{The valley instanton begins to deviate from the
instanton at $\rho v\lsim 1$.
As is shown later, the valley instanton at $\rho v=1$ is relevant
to the baryon number violating process with many gauge and Higgs
particles, $n_{_{\rm W}}+n_{_{\rm H}}\sim 40$.
Since the sphaleron decay, which is not a tunneling
process, becomes important at $n_{_{\rm W}}+n_{_{\rm H}}\lsim 40$
\cite{gold}, this deviation could be naturally explained.}
This suggest that the determinant in the background of the valley
instanton can be well-evaluated by using the instanton.
If the same approximation is used in the constrained instanton, the
errors are increased.

Finally we will discuss the implication to the baryon and lepton number
violating process in the standard model.
It is expected that the process at high energy is dominated by the one
containing many bosons in the process \cite{gold,Ring,Esp}.
Using the constrained instanton
with a small radius, the cross section of the process
${\rm q}+{\rm q}
\rightarrow7\bar{{\rm q}}+3\bar{l}
+n_{_{\rm W}}{\rm W}+n_{_{\rm H}}{\rm H}$,
where W stands for ${\rm W}^{\pm}$ and Z boson and H for Higgs boson,
was calculated in Refs.\cite{Ring,Esp}.
In their calculation, larger size instanton-like
configurations become important as $n_{_{\rm W}}$ and $n_{_{\rm H}}$
increase.
Since the authors of Refs.\cite{Ring,Esp} used the small size constrained
instanton, their calculations break down when the number of bosons is
large enough.
In section \ref{sec:GH}, we have revealed that the action of the constrained
instanton begins to deviate from the one obtained analytically by the
small size constrained instanton at $\rho v\sim 0.5$.
Namely $g^2\Delta S/8\pi^2=-0.0086$ at $\rho v=0.5$.
Since the action $S$ appears as $e^{-S}$ in the calculation, this gives
a significant difference in the amplitude.
In the standard model $g^2$ is given by $g^2\sim 0.42$ at the Z
boson mass scale. Therefore, the differences become $e^{-\Delta S}\sim 5$.
Using the result that the $\rho$ integral in the amplitude is dominated at
$\rho v=[(n_{_{\rm W}}+n_{_{\rm H}}+8.08)\,g^2(\mu)/2\pi^2]^{1/2}$
\cite{Esp},
it is found that the calculations in Refs.\cite{Ring,Esp} break down at
$n_{_{\rm W}}+n_{_{\rm H}}\sim 4$.

At the same time, the constrained instanton deviates from the valley
instanton.
Therefore at $n_{_{\rm W}}+n_{_{\rm H}}\lsim 4$, the valley instanton
is needed to obtain the plausible result of the amplitude.
If we adopt the valley instanton, it is possible to perform the
calculation of the amplitude including more gauge or Higgs bosons.
The most difficult part of the calculation is the evaluation of the
determinant resulted from the Gaussian integration.
However, as was mentioned above,
the determinant resulted from the Gaussian integration can be well-evaluated
in the valley instanton even when $\rho v=1$, which is relevant to the
calculation for $n_{_{\rm W}}+n_{_{\rm H}}\sim 40$.
Residue of the pole at $p=m_{_H}$ in the Higgs field and that at
$p=m_{_W}$ in $W^{\pm}$ and $Z$ fields, which are needed in the
calculation of the amplitude, can be calculated numerically.
As has been shown in subsection \ref{sec:fermion}, incorporation of the
light fermions into the SU(2) gauge-Higgs system is straightforward
even when the radius of the valley instanton is large.
For heavy fermion like top quark, the fermionic zero mode around the
valley instanton can be constructed numerically.
The analysis is in progress and will be
reported in the near future.

\vskip 1cm
\centerline{\large\bf Acknowledgment}

\noindent
We would like to thank our colleagues in Kyoto University for
useful discussions and encouragements.
One of the authors (H.A.) is supported in part by the
Grant-in-Aid \#C-07640391 from the Ministry of Education,
Science and Culture, Japan.
M.~S. and S.~W. are the fellows of the Japan Society for
the Promotion of Science for Japanese Junior Scientists.

\newpage

%%%%%  Definitions %%%%%%%%%%%%%%%%%%%%%%%%%%%%%%%%%%%%%%%%%%%%%%%%%%%%%
\newcommand{\J}[4]{{\sl #1} {\bf #2} (19#3) #4}
\newcommand{\MPL}{Mod.~Phys.~Lett.}
\newcommand{\NP}{Nucl.~Phys.}
\newcommand{\PL}{Phys.~Lett.}
\newcommand{\PR}{Phys.~Rev.}
\newcommand{\PRL}{Phys.~Rev.~Lett.}
\newcommand{\AP}{Ann.~Phys.}
\newcommand{\CMP}{Commun.~Math.~Phys.}
\newcommand{\CQG}{Class.~Quant.~Grav.}
\newcommand{\PRP}{Phys.~Rept.}
\newcommand{\SPU}{Sov.~Phys.~Usp.}
\newcommand{\RMPA}{Rev.~Math.~Pur.~et~Appl.}
\newcommand{\SPJ}{Sov.~Phys.~JETP}
\newcommand{\MP}{Int.~Mod.~Phys.}
%%%%  Contents %%%%%%%%%%%%%%%%%%%%%%%%%%%%%%%%%%%%%%%%%%%%%%%%%%%%%%%%%%

\end{document}